\documentclass[a4paper,11pt]{article}
\pdfoutput=1
\usepackage{jcappub}[=2021-04-14]
\arxivnumber{2101.02713}

\usepackage{bm}
\usepackage{bbold}

\def\segb{\sigma_{\rm BBH+BNS}}
\def\sugb{\sigma_{\rm UGB}}

\title{Measuring the primordial gravitational wave background in the presence of other stochastic signals}

\author{D. Poletti}
\emailAdd{davide.poletti@sissa.it}
\affiliation{International School for Advanced Studies (SISSA),\\
Via Bonomea 265, 34136, Trieste, Italy}
\affiliation{Institute for Fundamental Physics of the Universe (IFPU),\\
Via Beirut 2, 34014 Trieste, Italy}
\affiliation{National Institute for Nuclear Physics (INFN) Sezione di Trieste,\\
Padriciano, 99, 34149 Trieste, Italy}

\abstract{Standard methodologies for the extraction of the stochastic gravitational wave background (SGWB) from auto- or cross-correlation of interferometric signals often involve the use of a filter function.
The standard optimal filter maximizes the signal-to-noise ratio (SNR) between the total SGWB and the noise.
We derive expressions for the optimal filter and SNR in the presence of a target SGWB plus other unwanted components. We also generalize the methodology to the case of template-free reconstruction.
The formalism allows to easily perform analyses and forecasts that marginalize over foreground signals, such as the typical $\Omega_{\rm GW} \propto f^{2/3}$ background arising from binary coalescence.
We demonstrate the methodology with the LISA mission and discuss possible extensions and domains of~\mbox{application}.}
\providecommand\inspire[1]{\href{https://inspirehep.net/search?p=find+#1}{{\tiny IN}{\footnotesize SPIRE}}}

\providecommand\erratum[4][ibid.\ ]{\emph{Erratum #1}{\bf #2} (#3) #4}

\providecommand{\jhep}[3] {\ifnum#2>2009%
\href{https://doi.org/10.1007/JHEP#1(#2)#3}{\emph{JHEP} {\bf #1} (#2) #3}%
\else%
\href{https://doi.org/10.1088/1126-6708/#2/#1/#3}{\emph{JHEP} {\bf #1} (#2) #3}%
\fi}
\providecommand{\jcap}[3] {\href{https://doi.org/10.1088/1475-7516/#2/#1/#3}{\emph{JCAP} {\bf #1} (#2) #3}} 
\def\issueFromCounter.#1#2#3#4#5#6.{#2#3}
\providecommand{\jstat}[2]{\PackageWarningNoLine{\jname}{The macro \protect\jstat\space is obsolete!\MessageBreak Please typeset JSTAT as any other journal}%
  \href{https://doi.org/10.1088/1742-5468/#1/\issueFromCounter.#2./#2}{\emph{J.\ Stat.\ Mech.\ }(#1) #2}} 
\providecommand{\hepth}[1]{\href{https://arxiv.org/abs/hep-th/#1}{\tt hep-th/#1}}

\providecommand{\grqc}[1]{\href{https://arxiv.org/abs/gr-qc/#1}{\tt gr-qc/#1}}

\providecommand{\astroph}[1]{\href{https://arxiv.org/abs/astro-ph/#1}{\tt astro-ph/#1}}

\providecommand{\arXivid}[1]{\href{https://arxiv.org/abs/#1}{\tt arXiv:#1}}
\providecommand{\Math}[2]{%
\if!#1!%
\href{https://arxiv.org/abs/math/#2}{\tt math/#2}%
\else%
\href{https://arxiv.org/abs/math.#1/#2}{\tt math.#1/#2}%
\fi}

\begin{document}
\maketitle
\flushbottom

\section{Introduction}
\label{sec:intro}
Measuring the stochastic gravitational wave background (SGWB) is among the main objectives of existing and planned gravitational wave observatories.
It is expected to be produced by a wide variety of astrophysical or cosmological phenomena, see~\cite{Maggiore:2000gv,Christensen:2018iqi,Caprini:2018mtu,Regimbau:2011rp} for reviews.
An astrophysical SGWB certainly exists.
In recent years, we have witnessed a series of detections of gravitational waves transient signals produced by binary back holes and binary neutron star mergers~\cite{Abbott:2016nmj, Abbott:2016blz, TheLIGOScientific:2016pea, Abbott:2017vtc, Abbott:2017gyy, Abbott:2017oio, TheLIGOScientific:2017qsa}.
The signal produced by these systems, integrated over the whole history of the universe, must constitute a background of gravitational waves.
Other contributions are expected to come from core-collapse supernovae, non-axisymmetric spinning neutron stars and magnetars (see, e.g.,~\cite{Buonanno:2004tp,Yakunin:2010fn,Ferrari:1998jf,Cheng:2015rja}).
The background of gravitational waves produced through these mechanisms offers a window on a large number of astrophysical processes over the entire history of the universe.

The SGWB is also considered a major probe of inflation~\cite{Starobinsky:1980te,Linde:1981mu,Guth:1980zm,Mukhanov:1981xt,Bartolo:2016ami,Guzzetti:2016mkm}.
By postulating a period of accelerated expansion before the standard Big-Bang universe, inflation is capable of solving the horizon, flatness and monopole problems.
The standard inflationary mechanism typically involves a particle beyond the Standard Model dominating the energy content of the infant universe.
Its quantum fluctuations --- blown up by the quasi-De Sitter expansion --- naturally act as seeds of the cosmological structures we observe in the low redshift universe as well as the acoustic waves observed in the cosmic microwave background (CMB) at $z \simeq 1100$.
Quantum fluctuations of the gravitational field itself are expected to undergo a similar fate and additional GW contributions may arise from the presence of fields in addition to or in place of the simple scalar-field inflaton. The resulting SGWB should be still present today at the frequencies of existing and planned GW direct detection experiments.
Likewise, these GW would leave an imprint in the CMB, best observable in the sought-after B-mode polarization, a key target of the ultimate (or near ultimate) CMB observatories~\cite{Hazumi:2019lys,Abazajian:2019eic}.

Topological strings~\cite{Kibble:1976sj,Vachaspati:2015cma} and superstrings~\cite{Sarangi:2002yt,Jones:2003da} attracted a lot of attention in this context. They also generate an SGWB resulting from the superposition of powerful bursts of GWs produced by cusps and kinks propagating on string loops.
Again, this type of stochastic background can be constrained with both CMB~\cite{Pagano:2015hma} and direct detection experiments~\cite{Abbott:2017mem}.

Either cosmological or astrophysical, no detection of SGWB was obtained so far. Nevertheless, the measurements of individual gravitational wave events~\cite{Abbott:2016nmj, Abbott:2016blz, TheLIGOScientific:2016pea, Abbott:2017vtc, Abbott:2017gyy, Abbott:2017oio, TheLIGOScientific:2017qsa} allowed to refine the predicted amplitude of the SGWB from compact binary coalescences~\cite{TheLIGOScientific:2016wyq,Abbott:2017xzg} at 25 Hz --- the frequency where the aLIGO and aVIRGO observatories are most sensitive to the SGWB\@.
Moreover, at the other end of the frequency spectrum --- in the 1--100 nHz range --- the common-spectrum stochastic signal detected by the NANOGrav Collaboration across an array of 45 pulsars~\cite{Arzoumanian:2020vkk} may be sourced by an SGWB, even though we have to wait for conclusive evidence of the quadrupolar correlations~\cite{Hellings:1983fr} before claiming detection of SGWB and making inferences on its origin.

It is clear that some form of SGWB exists and, given the wide variety of possible origins, the analyses (as well as the forecasts) will have to account for the possibility that multiple background sources are contributing at the same time.
Some of them will constitute the target SGWB, while the others will be considered as foregrounds to be removed.
The problem has been already explored~\cite[see, e.g.,][]{Pan:2019uyn,Flauger:2020qyi,Pieroni:2020rob} and
it was recognized that under some conditions the SGWB extraction and foreground rejection reduces to a linear problem~\cite{Parida:2015fma}.

In this paper, we show that, under the same conditions, the foregrounds can be rejected without explicitly estimating their amplitude, as opposed to what is done in the existing literature.
After briefly summarizing the main quantities involved in the studies of SGWBs (section~\ref{s:sgwb}), we review the standard approach~\cite{Allen:1996vm,Allen:1997ad} to maximize the SNR in foreground-free data (section~\ref{s:std_filt}).
It is based on the correlation of interferometry signals employing a frequency-dependent filter, which determines the optimality of the correlation.
In section~\ref{s:new_filt} we present a new filter that, unlike the standard one, can account for the presence of foreground signals and optimally marginalizes over them, possibly exploiting external prior information (see appendix~\ref{s:new_filt_dem} for the demonstration).
The expressions of both the new filter and the corresponding SNR have the attractive property of being nearly as simple as the standard ones while all the additional terms --- arising from the marginalization over the foregrounds --- have an intuitive and clear interpretation.
In section~\ref{s:template_free} and appendix~\ref{s:fisher} we generalize the approach to allow for the reconstruction of SGWBs that are not known a priori.
Also in this case, the main difference with respect to the existing literature is that we marginalize over the foregrounds without explicitly estimating their amplitudes.
In section~\ref{s:application} we show an application of both the new filter and the template-free reconstruction to the LISA mission. We summarize our results and conclude in section~\ref{s:conclusions}.

\section{The stochastic backgrounds of gravitational waves}
\label{s:sgwb}
We summarize here the key definitions about isotropic and unpolarized SGWB, to which we restrict ourselves in the present paper. The background of gravitational waves can be expressed as a plane wave expansion of the metric perturbation in the transverse-traceless gauge as
\begin{equation}
h_{ab}(t, \vec x) = \sum_P \int_{-\infty}^\infty df\ \int_{S^2}
d^2\hat n\ h_P(f,\hat n)\ e^{i2\pi f(t-\hat n \cdot\vec x/c)}\
e_{ab}^P(\hat n)\,,
\label{e:h_ab}
\end{equation}
where the wave vector has been expressed in terms of its normalization and
the unit vector specifying the direction of propagation, $\vec k \equiv 2 \pi f \hat n / c$.
We project the polarization state of the gravitational wave on the ``plus'' and ``cross'' bases, $P = +, \times$,
\begin{align}
e_{ab}^+(\hat n) & = \hat{\mathbb{m}}_a \hat{\mathbb{m}}_b - \hat{\mathbb{n}}_a \hat{\mathbb{n}}_b\,,
\label{e:e_ab^+}\\
e_{ab}^\times(\hat n) & =  \hat{\mathbb{m}}_a \hat{\mathbb{m}}_b + \hat{\mathbb{n}}_a \hat{\mathbb{n}}_b\,,
\label{e:e_ab^x}
\end{align}
where $\hat{\mathbb{m}}$ and $\hat{\mathbb{n}}$ are unit vectors orthogonal to $\hat  n$ and to each other. Note that the basis is normalized such that $e^P_{ab} e^{P', ab} = 2 \delta^{PP'}$.
In this article we assume that the observed gravitational waves belong to an SGWB that is Gaussian distributed, with zero mean and~variance
\begin{equation}
\langle h_P^*(f, \hat  n) h_{P'}(f',\hat  n')\rangle = \frac{1}{8 \pi}
\delta^2(\hat  n ,\hat  n')\delta_{PP'}\delta(f-f')\ S_h(f)\,,
\label{e:S_h}
\end{equation}
where the delta functions reflect the assumption that the background is isotropic, unpolarized and stationary.
The spectral density function $S_h(f)$ is a real function satisfying $S_h(-f) = S_h(f)$.
Eq.~(\ref{e:S_h}) also sets our convention for its normalization, which varies across the literature and in our case satisfies
\begin{equation}
\sum_{P\ P'} \int d\hat n\  d\hat n'\
\langle h_P^*(f, \hat  n) h_{P'}(f',\hat  n')\rangle =
\delta(f-f')S_h(f)\,.
\label{e:S_h_avg}
\end{equation}
The information carried by $S_h$ is often expressed in terms of
\begin{equation}
\Omega_{\rm gw}(f) \equiv \frac{1}{\rho_{\rm cr}}\
\frac{d\rho_{\rm gw}(f)}{d\ln f}\,,
\label{e:Omega_gw}
\end{equation}
where $\rho_{\rm cr} = 3 c^2 H_0^2 / 8 \pi G$ it the critical density of the universe today and
\begin{equation}
\rho_{\rm gw} = \frac{c^2}{32 \pi G}  \langle \dot h_{ab}(t,\vec x) \dot h^{ab}(t,\vec x) \rangle
\label{e:rho_gw}
\end{equation}
is the energy density in gravitational waves.
Given our convention, $\Omega_{\rm gw}$ and $S_h$ are related~by
\begin{equation}
\Omega_{\rm gw}(f) = \frac{4 \pi^2}{3 H_0^2} f^3 S_h(f)
\label{e:O_gw}
\end{equation}
\section{Measuring the stochastic background}
\label{s:measuring}
The output of a gravitational wave detector $I$ is a time stream
\begin{equation}
d_I(t) = s_I(t) + n_I(t)\,.
\label{e:d_t}
\end{equation}
The signal $s$ is a real function of time and is  sourced by the gravitational waves hitting the detector
\begin{equation}
s_I(t) = \sum_{P = +, \times } \int_{-\infty}^{\infty} d f \int d^2\hat n\
F_I^P(\hat n, f)\  h_P(f,\hat n)\ e^{i2\pi f(t-\hat n \cdot\vec x_I/c)}
\,,
\label{e:s_t}
\end{equation}
where  $\vec x$ is the location of the detector and the response function $F^P_I(\hat n,  f)$ depends on properties of the detector such as the geometry and orientation --- which we assume constant in time.
The noise term $n_I$ is also a real quantity, assumed to be stationary and Gaussian, with variance given by
\begin{equation}
\langle \tilde n_I(f) \tilde n_J^*(f') \rangle = \frac{1}{2} \delta (f - f') N_{IJ}(f)\,.
\label{e:N_IJ}
\end{equation}
The $1/2$ prefactor is  conventional but, unlike the one of $S_h$, this normalization is consistent across the literature. Note that in eq.~(\ref{e:N_IJ}) we have considered the possibility that $I$ and $J$ are  different detectors with correlated noise, as in the case of the $XYZ$ channels of the LISA mission~\cite{2012CQGra..29l4015V}.

The searches of stochastic gravitational wave backgrounds typically involve the correlation of (possibly different) detectors, $I$ and $J$, both collecting signal over a time $T$
\begin{equation}
x_{IJ} \equiv \int_{-T/2}^{T/2} dt\ \int_{-T/2}^{T/2} dt'
\left( d_I(t)\ d_J(t') - \frac{1}{2} N_{IJ}(|t - t'|) \right)\ Q(t, t')\,.
\label{e:cross_t}
\end{equation}
The second term in the parenthesis subtracts the noise bias when the noise in $I$ and $J$ is correlated --- or they are in fact the same detector.\footnote{Note that this second term is missing in the large portion of literature that focuses on uncorrelated detectors, in which case $N_{IJ}(f) \propto \delta_{IJ}$.}
It is related to eq.~(\ref{e:N_IJ}) by
$N_{IJ}(|t - t'|) / 2 = \langle n_I(t) n_J(t') \rangle = \int_{-\infty}^{\infty} df e^{i 2 \pi f(t - t')} N_{IJ}(f)/2$.
The filter function $Q$ can be arbitrarily chosen in order to maximize the signal-to-noise of the cross-correlation.
In section~\ref{s:std_filt} and~\ref{s:new_filt}, we discuss in detail the optimization of $Q$.
For time being, we only note that because of the stationarity of both signal and noise, $Q$ must be a function of $|t - t'|$.
We assume now that $Q(|t - t'|)$ is non-negligible only for time differences much smaller than the observation time.
This allows to push to infinity the extremes of one of the integrals in eq.~(\ref{e:cross_t}), and thus to~get\footnote{The tilde denotes the Fourier transform, defined as $\tilde g(f) \equiv \int_{-\infty}^\infty dt\ e^{-2\pi f t i} g(t)$.}
\begin{equation}
x_{IJ} = \int_{-\infty}^{\infty} df\int_{-\infty}^{\infty} df'
\ \delta_T(f-f') \tilde d_I^*(f)\ \tilde d_J(f') \tilde Q(f') -  \frac{T}{2} \int_{-\infty}^{\infty} df N_{IJ}(f) \tilde Q(f)\,.
\label{e:cross_f}
\end{equation}
The function $\delta_T(f) = \sin(\pi f T) / \pi f$ has the following properties.
It converges to the Dirac $\delta$ function as $T \to \infty$.
In the same limit, $\delta_T^2$ converges to $T \delta$.
For finite $T$, $\delta_T(0) = T$.

The expected value and variance of $x$ can be computed from those of $\tilde d_I^*(f)\ \tilde d_J(f')$, which in turn have simple expressions thanks to eqs.~(\ref{e:S_h_avg}),~(\ref{e:s_t}) and~(\ref{e:N_IJ})---assuming that all the non-stationary, anisotropic or polarized signals have been removed or are negligible---,
\begin{equation}
\langle x_{IJ} \rangle = T \int_{0}^{\infty} df S_h(f) \mathcal{R}_{IJ}(f) \tilde Q(f)\,,
\label{e:e_cross}
\end{equation}
where we have defined the response function
\begin{equation}
\mathcal{R}_{IJ} (f) \equiv \frac{1}{4 \pi} \int_{S^2} d^2 \hat n \sum_P F^{*P}_I(\hat n, f)  F^P_J(\hat n, f) e^{-i 2 \pi f \hat n \cdot (\vec{x}_J - \vec{x}_i) / c},
\end{equation}
which depends on the properties of the detectors and their relative distance and orientation.

The variance receives a contribution from both the signal and noise, but the noise power is typically much higher than the signal.
Therefore, we can ignore $s$ in the time streams of the detectors and get
\begin{equation}
\langle x_{IJ}^2 \rangle - \langle x_{IJ} \rangle^2 =
\frac{T}{2} \int_{0}^{\infty} df \left(N_{II} (f) N_{JJ}(f) + N_{IJ}^2(f)\right)| \tilde Q (f)|^2 \,.
\label{e:var_cross}
\end{equation}
The following two sections discuss how to optimize the signal-to-noise ratio.
\begin{equation}
{\rm SNR} = \frac{\langle x_{IJ} \rangle}{\sqrt{\langle x_{IJ}^2 \rangle - \langle x_{IJ} \rangle^2}}
\end{equation}
\subsection{Standard optimal filtering}
\label{s:std_filt}
In this section we derive the standard expression of the optimal $Q$~\cite{Allen:1996vm,Allen:1997ad}, thus summarizing the approach commonly adopted in the literature and paving the road for the new filter that we propose in the next section.

We want to find the $Q(f)$ that maximizes the SNR\@.
In order to do it, it is very convenient to express eqs.~(\ref{e:e_cross}) and~(\ref{e:var_cross}) in term of the following inner product of complex functions
\begin{equation}
\bm{a} \cdot \bm{b} = \frac{T}{2}\int_{0}^{\infty} df a^*(f) b(f) N^2(f)
\end{equation}
where we have defined for convenience
\begin{equation}
N^2(f) \equiv N_{II} (f)\ N_{JJ}(f) + N_{IJ}^2(f)\,.
\end{equation}
In contrast with the entire existing literature, we include $T$ in the definition of the inner product.
We denote with boldface the objects (lowercase for vectors and upper case for matrices) that obey to this inner product.
For convenience we define the following vectors
\begin{align}
\bm{h} & \equiv  \frac{2\ S_h(f) \mathcal{R}_{IJ}(f)}{N^2(f)}
\label{e:vec_h}\\
\bm{q} & \equiv  \tilde Q (f)
\end{align}
The expression for the SNR becomes
\begin{equation}
    {\rm SNR}^2 = \frac{(\bm{q} \cdot \bm{h})^2}{\bm{q} \cdot \bm{q}}\,,
\end{equation}
which is obviously independent of the normalization of $\bm{q}$ and is maximized when the vector $\bm{q}$ is aligned to $\bm{h}$.
The optimal filter is thus
\begin{equation}
    \tilde Q (f) \propto \frac{\mathcal{R}_{IJ}(f) S_h(f)}{N^2(f)}
    \label{e:old_q}
\end{equation}
and the signal-to-noise achieved is
\begin{equation}
    {\rm SNR}^2 =
    2 T \int_{0}^{\infty} d f
    \frac{|\mathcal{R}_{IJ}(f)|^2 S^2_h(f)}{N^2(f)} \,.
    \label{e:old_snr}
\end{equation}
Other expressions in the literature do not have the factor 2. This is either because their integral ranges from $-\infty$ to $\infty$ or due to the fact they are considering an auto-correlation of a channel $I$, in which case $N^2(f) = 2 N^2_{II}(f)$.

\begin{sloppypar}
\subsection{A new generalized filter: accounting for unwanted components in the SGWB}
\label{s:new_filt}
The filter presented in the previous section has one limitation: it can only optimize the detection of a single, global gravitational wave background.
In particular, it can not accommodate for the presence of multiple components with unknown relative amplitudes, or with amplitudes known up to some uncertainty.
We now propose a new filter capable of handling this generalized SGWB.
\end{sloppypar}

We study explicitly the case of a two-components SGWB
\begin{equation}
    S_h(f) = S_p(f) + \alpha S_{a}(f),
    \label{e:S_h_model}
\end{equation}
where we want to optimize the filter $\tilde{Q}(f)$ for the measurement of  a primordial SGWB $S_p(f)$, in the presence of an astrophysical contribution with known shape $S_a(f)$ and unknown amplitude $\alpha$, which is the only free parameter of the model.
We comment on models with non-linear free parameters in appendix~\ref{s:non_lin_model}.
We allow for some external information on $\alpha$ to be available.
We indeed assume its expected value to be $\bar{\alpha}$ and its variance $\sigma^2$.
The lack of such external information is represented by the limit $\sigma\to\infty$, in which case the chosen value of $\bar{\alpha}$ becomes irrelevant.

It is natural to amend the cross-correlation estimator to remove the expected contribution from astrophysical sources
\begin{equation}
    y_{IJ} \equiv x_{IJ} - \bar \alpha\ T \int_0^{\infty} df\ S_a(f) \mathcal{R}_{IJ}(f) \tilde Q(f)\,.
    \label{e:y_cross}
\end{equation}
As in the previous section, we should now maximize the SNR for this new cross-correlation estimator, taking into account that the amplitude of the astrophysical signal we subtracted is uncertain.
Note that the noise bias removed in eq.~(\ref{e:cross_f}) is not any different from the removal of an astrophysical component with $\sigma = 0$.

In this more general data model, the optimal filter becomes
\begin{equation}
    \tilde Q (f) \propto
    \frac{\mathcal{R}_{IJ}(f) S_p(f)}{N^2(f)}
    -
    \frac{\mathcal{R}_{IJ}(f) S_a(f)}{N^2(f)}
    \frac{2 T\int_{0}^{\infty} d f'\ |\mathcal{R}_{IJ}(f')|^2\ S_a(f') S_p(f') N^{-2}(f')}
    {\sigma^{-2} + 2 T \int_{0}^{\infty} d f'\ |\mathcal{R}_{IJ}(f')|^2\ S^2_a(f') N^{-2}(f')},
    \label{e:new_q}
\end{equation}
and the corresponding SNR is
\begin{equation}
    {\rm SNR}^2 =
    2 T \int_{0}^{\infty} d f
    \frac{|\mathcal{R}_{IJ}(f)|^2 S^2_p(f)}{N^2(f)}
    -
    \frac{
    \left[
    2 T\int_{0}^{\infty} d f
    |\mathcal{R}_{IJ}(f)|^2 S_a(f)S_p(f)N^{-2}(f)
    \right]^2
    }
    {\sigma^{-2}
     +
     2 T\int_{0}^{\infty} d f
     |\mathcal{R}_{IJ}(f)|^2 S^2_a(f)N^{-2}(f)
    }.
    \label{e:new_snr}
\end{equation}
These two expressions are the main result of this paper. We report their derivation in appendix~\ref{s:new_filt_dem}, where we also consider the more general case of an arbitrary number of contributions to $S_h(f)$.

In both eq.~(\ref{e:new_q}) and~(\ref{e:new_snr}), the first term corresponds to the standard filter and SNR, reported in eqs.~(\ref{e:old_q}) and~(\ref{e:old_snr}).
The second term is the correction that accounts for the presence of the astrophysical component in the SGWB\@.
Note that in eq.~(\ref{e:new_snr}) this term is always negative, reflecting the intuition that the presence of astrophysical sources can only degrade the SNR.

Looking at the numerator more closely, the integral (and thus the SNR degradation) is maximum when the primordial and astrophysical components have the same frequency dependence.
When this happens the SNR becomes
\begin{equation}
    {\rm SNR}^2 =
    \frac{
    2 T\int_{0}^{\infty} d f
    |\mathcal{R}_{IJ}(f)|^2 S^2_p(f)N^{-2}(f)
    }
    {
    1 + \sigma^2
     2 T\int_{0}^{\infty} d f
     |\mathcal{R}_{IJ}(f)|^2 S^2_a(f)N^{-2}(f)
    }
    \ \ \ \ {\rm for }\ \ \ \ S_a(f) \propto S_p(f),
\end{equation}
which is zero for $\sigma\to\infty$: when the primordial and astrophysical SGWB have the same frequency dependence, nothing can be said on the primordial component if no external information on the astrophysical one is available.

Focusing now on the denominator of the second term in eq.~(\ref{e:new_snr}), it is the inverse-variance on the estimation of $\alpha$ if no primordial signal was present.
If both $\alpha$ and the noise are Gaussian, it represents the total Fisher information about the amplitude of the astrophysical SGWB\@.
It is clearly separated into the external and the internal contribution, with the latter always overtaking the former for sufficiently long observational times.

\section{Template-free reconstruction}
\label{s:template_free}
Building on the approach illustrated in the previous section, we now extend the discussion to the reconstruction of the primordial SGWB without assuming a template for it.
More details are provided in appendix~\ref{s:fisher}, together with the general expression for an arbitrary number of foreground components.

We start from the following quantity,
\begin{equation}
    z_{IJ}(f) \equiv \frac{2}{T \mathcal{R}_{IJ}(f)} \int_0^{\infty}d f' \left(d_I^*(f)d_J(f') + d_I(f)d_J^*(f')\right) \delta_T(f-f') - \frac{N_{IJ}(f)}{\mathcal{R}_{IJ}(f)} - \bar{\alpha } S_a(f).
    \label{e:z}
\end{equation}
It is essentially an elaboration on eq.~(\ref{e:cross_f}), without the integration over $f$, and represents a template-free reconstruction with the highest resolution allowed by the observations $I$ and $J$ --- the expected value is indeed $\langle z_{IJ}(f) \rangle = S_p(f)$.
The two terms inside the integral (instead of one) have the only effect of folding the integral, which ranges over positive values of $f'$.
All the terms outside of the integral just remove additive and multiplicative biases.
The variance of $z_{IJ}$ is
\begin{equation}
    V(f, f') = \frac{N^2(f)\delta(f - f')}{2 T |\mathcal{R}_{IJ}(f)|^2} +\sigma^2\ S_{a}(f)  S_{a}(f')
    \label{e:z_var}
\end{equation}
and its inverse is equal to
\begin{equation}
    F(f, f') = \frac{2 T|\mathcal{R}_{IJ}(f)|^2}{ N^{2}(f)} \delta(f - f') -
    \frac{
    \left[ 2 T  |\mathcal{R}_{IJ}(f)|^2 S_a(f)N^{-2}(f) \right]
    \left[ 2 T  |\mathcal{R}_{IJ}(f')|^2 S_a(f')N^{-2}(f') \right]
    }
    {
    \sigma^{-2}
     +
     2 T \int_{0}^{\infty} d f''
     |\mathcal{R}_{IJ}(f'')|^2 S^2_a(f'')N^{-2}(f'')
    }
    \,.
    \label{e:fisher}
\end{equation}
In the Gaussian approximation $F$ coincides with the Fisher matrix, the information available for any SGWB model for the given experimental configuration. In any case, it is related to the SNR on a specific model in  eq.~(\ref{e:new_snr}) by
\begin{equation}
 {\rm SNR}^2 = \int_0^\infty df \int_0^\infty df' F(f, f')  S_p(f') S_p(f) \,.
 \label{e:snr_info}
\end{equation}

$z_{IJ}$ is typically very noisy and it should be used as an intermediate step towards its projection onto a subspace defined by a basis of model spectra.
Using eq.~(\ref{e:fisher}) in the projection, we obtain a minimum variance estimation that takes into account both the propagation of the instrumental noise at each frequency and the foreground contamination due to an imperfect (i.e., noisy) estimation of $\alpha$.
The key point in our approach is that $\alpha$ is not explicitly estimated but, if it was, the result would be statistically equivalent.
This latter solution consists of including the astrophysical template(s) in the model spectra and applying the technique illustrated by~\cite{Parida:2015fma}, which uses only the first term in eq.~(\ref{e:fisher}) to define the projection.
The two approaches are largely equivalent, in appendix~\ref{s:fisher} we argue that ours have some potential numerical advantage, mainly in the case of several near-degenerate astrophysical components.

The most important aspect in the projection of $z_{IJ}$ onto a basis of models is anyway the choice of the models themselves.
They can be motivated by the theory (e.g., SGWB spectra from a class of physically motivated models), by a target property of the reconstructed spectrum (e.g., polynomials of varying nature or some type of smooth functions to avoid sharp features) or by the constraints that the experiment can provide over them.
The principal component analysis (PCA) is part of this last category and its usage was already proposed by~\cite{Pieroni:2020rob}, even though in appendix~\ref{s:fisher} we argue that its interest in this context is limited.

\section{An example application: the LISA mission}
\label{s:application}
In this section we show an example of how the formalism of the previous section can be applied. We forecast the performance of the LISA mission attempting to detect a specific model of primordial SGWB in the presence of an astrophysical signal.

\subsection{The LISA experiment}
\begin{figure}
    \centering
    \includegraphics[width=0.32\linewidth]{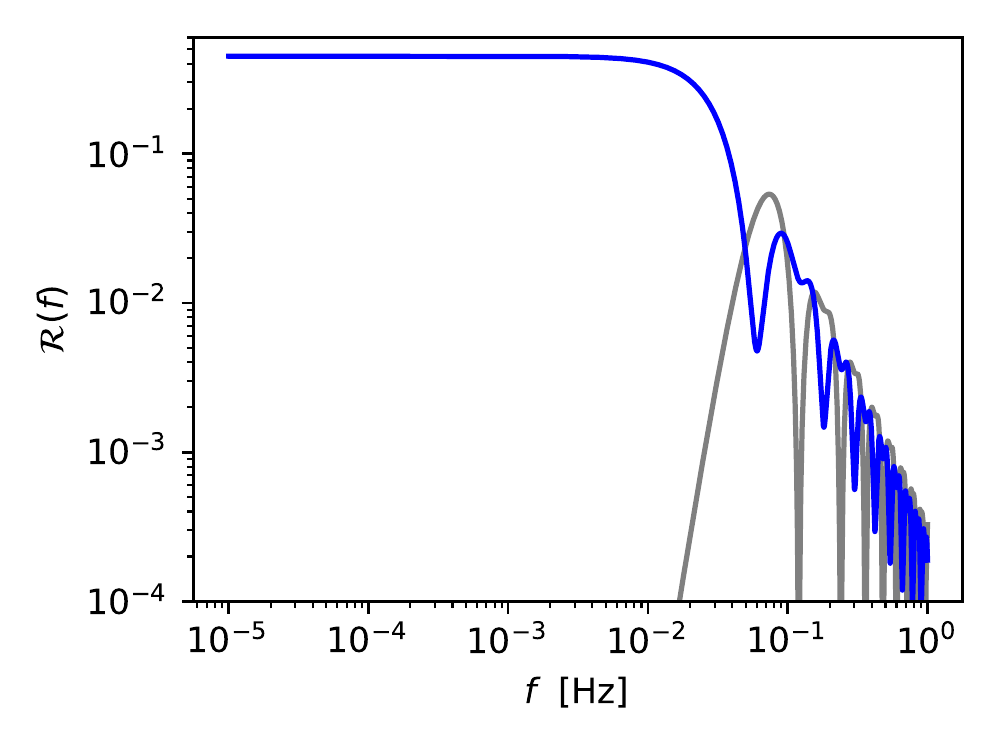}
    \includegraphics[width=0.32\linewidth]{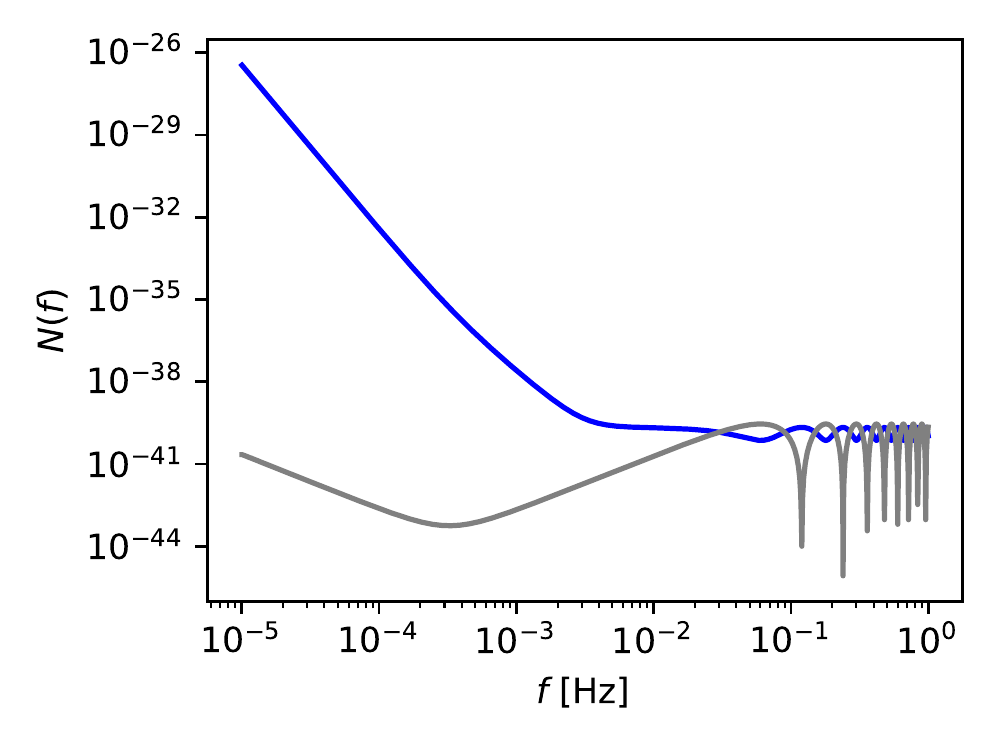}
    \includegraphics[width=0.32\linewidth]{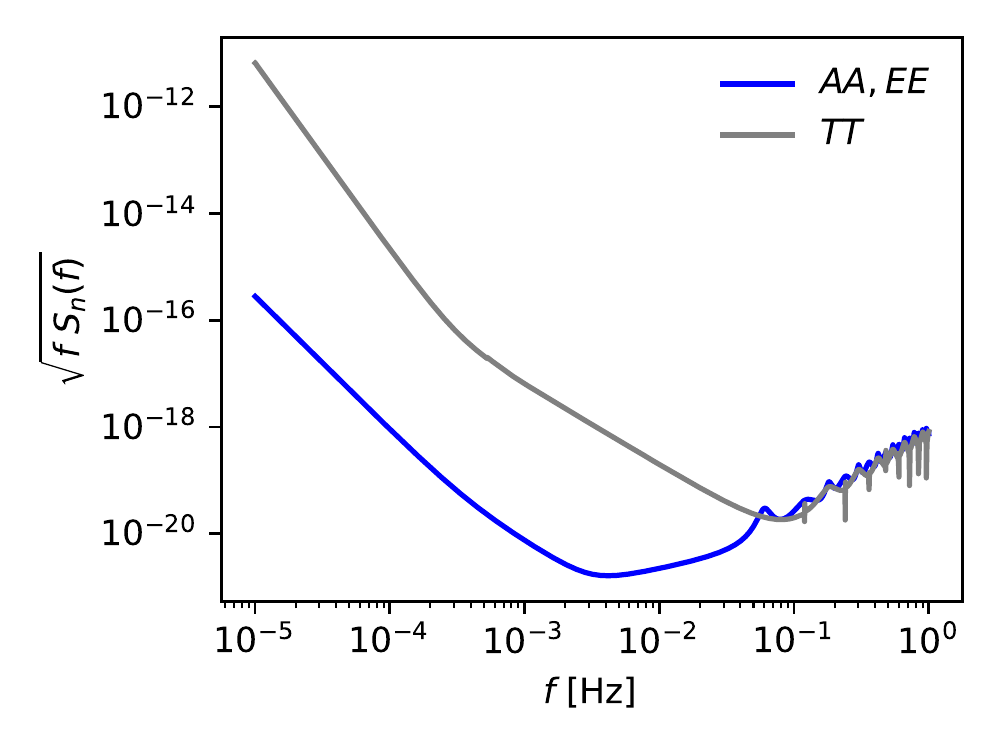}
    \caption{The main properties of the $A$, $E$ and $T$ channels of LISA, the experimental configuration considered in the demonstration in section~\ref{s:application}. \emph{Left:} response function.
    \emph{Center:} noise power spectral density.
    \emph{Right:} characteristic strain~sensitivity.\label{f:lisa}}
\end{figure}

The Laser Interferometer Space Antenna (LISA)~\cite{2017arXiv170200786A} is the most advanced project for a gravitational wave antenna in space.
It is an ESA L-class mission with a NASA partnership and it is planned for launch in the early/mid 2030s.
The experiment consists of three spacecraft, 2.5 million km apart, in a triangular configuration.
Laser links between the spacecrafts will provide three interferometry measurements which will monitor the relative distance between the test masses in the spacecrafts, allowing to probe the space-time distortion due to incoming gravitational waves.
Each of the three pairs of arms provides an interferometry measurement --- a time series as the one in eq.~(\ref{e:d_t}).
These are the so-called $XYZ$ channels. These signals are correlated but, thanks to the symmetry of the configuration, it is easy and natural to extract their eigenmodes, dubbed $AET$ channels, given by
\begin{align}
    d_A & =  (d_X - 2 d_Y + d_Z) / \sqrt{6}\\
    d_E & =  (d_X - d_Z) / \sqrt{2}\\
    d_T & =  (d_X + d_Y + d_Z) / \sqrt{3}.
\end{align}
While $A$ and $E$ have identical noise properties and response functions, those of the $T$ channel are very different and make this latter channel much less sensitive than the former ones (see figure~\ref{f:lisa}).
We also convert the $A/E$ noise power spectral density to energy density $\Omega_{\rm GW}$ with eq.~(\ref{e:O_gw}) and display it in figure~\ref{f:spectra_and_q}.
The calculation of the noise curves as well as the response functions\footnote{For the response functions we use the tabulated numerical values that the authors made available at \url{https://doi.org/10.5281/zenodo.3341817}.} follows~\cite{Smith:2019wny} and makes the same simplifying assumptions. For example, we assume that there is no gap in the data, the noise is perfectly stationary and we ignore the time dependence of the response function due to the orbital motion of the spacecrafts. We refer the reader to this article for more details.

\subsection{The SGWB}
We consider two types of backgrounds of astrophysical origin. The first one is sourced by the incoherent superposition of the emission of compact binaries that are not individually resolved by the experiment --- mostly stellar-origin black holes and neutron star binaries (BBH+BNS)\@.
This signal is well approximated by a power law~\cite{Regimbau:2011rp}
\begin{equation}
\Omega_{\rm GW} (f) h^2 = \Omega_* \left( \frac{f}{f_*} \right)^{2/3}, \ \ \ \ \ \ \ \  (BBH + BNS)
\end{equation}
We use $\Omega_* = 8.9 \times 10^{-10}$ at $f_* = 25$ Hz~\cite{LIGOScientific:2019vic} but we consider it an unknown parameter to marginalize over.

The second type of astrophysical background that we consider is sourced by unresolved galactic binaries (UGB)\@. We model its contribution with~\cite{Cornish:2017vip}
\begin{equation}
S(f) = A \left(\frac{1\ {\rm Hz}}{f}\right)^{7/3} \exp\left[ - \left(\frac{f}{1\ {\rm Hz}} \right)^\alpha - \beta f \sin(\kappa f) \right][1 + \tanh(\gamma (f_k - f))], \ \ \ \ \ \ \ \  (UGB)
\end{equation}
which we convert to $\Omega_{\rm GW}$ with eq.~(\ref{e:O_gw}). We fix the free parameters to the following values, which refer to a 4 year-long LISA mission, $\alpha = 0.138$,  $\beta = -221\ {\rm Hz}^{-1}$,  $\kappa = 521\ {\rm Hz}^{-1}$,  $\gamma = 1680\ {\rm Hz}^{-1}$. $f_k = 0.00113\ {\rm Hz}^{-1}$ and $A = 9 \times 10^{-45}\ {\rm Hz}^{-1}$~\cite{Cornish:2018dyw,Schmitz:2020rag}.

Note that both $A$ and $\Omega_*$ are included in the definition of the $\Omega_{\rm GW}$ of both UGB and BBH+BNS\@.
They are not absorbed by the normalization factor $\alpha$ of the respective components.
Therefore, we will assume the normalization factors to have priors centered at 1 and we will call $\sigma_{\rm BBH+BNS}$ and $\sigma_{\rm UGB}$ their standard deviations, which thus correspond to the relative uncertainty on $A$ and $\Omega_*$ from some external measurement.
We will swipe wide intervals of $\segb$ and $\sugb$, without necessarily justifying their values. The reason is that in this paper we do not aim at providing accurate forecasts but rather showing how the outcome of our methodology depends on its ingredients.

Finally, the primordial SGWB that we consider is the AX1 model of~\cite{Campeti:2020xwn}---which is produced by a spectator axion-SU(2) model.
The origin and properties of such a background are, however, irrelevant for our analysis.
We choose this model only for illustration purposes and are motivated mostly by its amplitude, which makes such a background within reach for the instrumental configuration of our choice.

We display the three components of the measured background in figure~\ref{f:spectra_and_q}.
Their different frequency dependence is the main feature that our methodology exploits, while their amplitudes constitute the free parameters of the model.

\subsection{Results for the A channel}
We now apply the formalism of section~\ref{s:new_filt} to the LISA $A$ channel and study the new filter and SNR\@. The results for the $E$ channel would be of course identical, while the $T$ channel is noise-dominated and we do not report results about it.

We consider values of $\sugb$ ranging from $10^{-4}$ and $10^{-2}$ and values of $\segb$ ranging from $10^{-3}$ and $10^{-1}$.
As we will see, the lower values correspond to the case where the external information completely constrains the astrophysical component, the higher value is equivalent to the case where no external information is available.
We stress that the range of values for $\sugb$ and $\segb$ is chosen only to best illustrate the phenomenological properties of our formalism and does not stem from an instrumental or theoretical forecast.
Also the inclusion of this level of UGB is not fully realistic, as this signal can be removed exploiting its time dependence owing to the motion of the spacecrafts~\cite{Adams:2013qma}.

\begin{figure}
    \centering
    \includegraphics[width=0.33\linewidth]{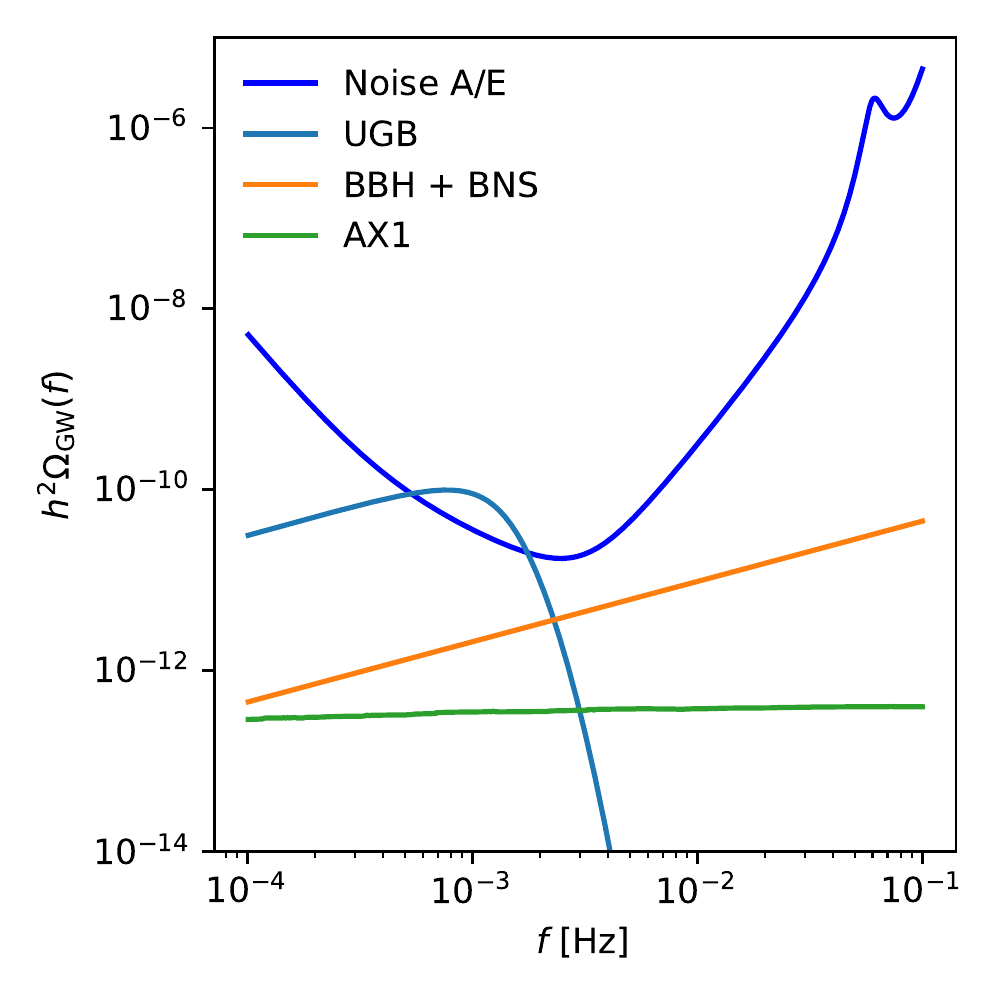}
    \includegraphics[width=0.66\linewidth]{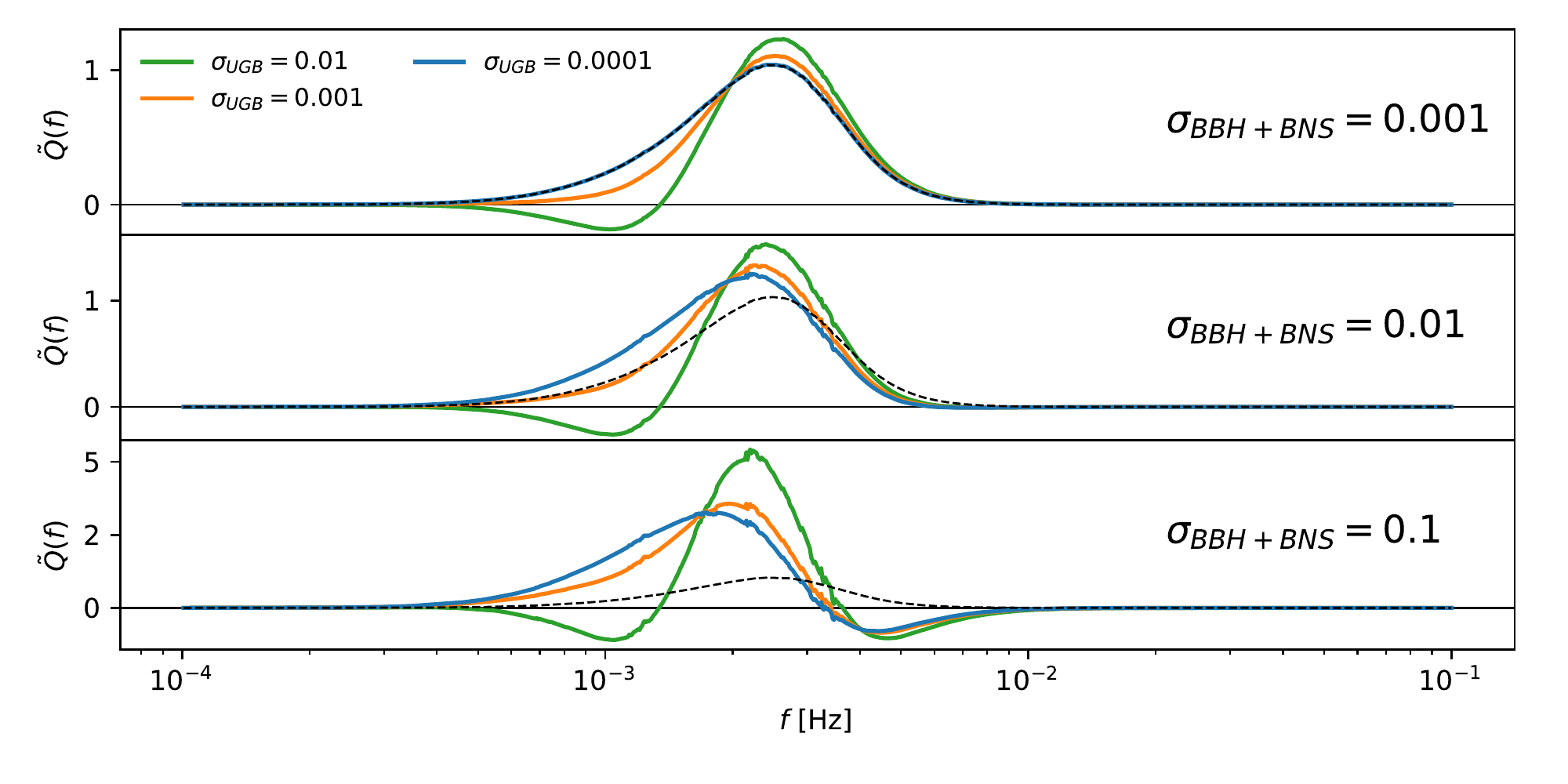}
    \caption{\emph{Left:} components of the SGWB considered in the application of the methodology. The noise equivalent $\Omega_{\rm GW}$ for the $A$ (or $E$) channel is also reported.
    \emph{Right:} shape of the filter functions for different values of $\sugb$ (color) and $\segb$ (subplots). For comparison, the standard filter is the dashed black~line.\label{f:spectra_and_q}}
\end{figure}

\paragraph{The filter.}
We start by focusing on the effect that the astrophysical components have on the filter eq.~(\ref{e:new_q})---and eq.~(\ref{e:new_q_vec}).
The three panels of figure~\ref{f:spectra_and_q} refer to three different amounts of external information on the extra-galactic binaries, while the three colors represent different priors on the galactic binaries.
The dashed line reports the standard filter, which only performs inverse noise co-addition: it is always positive and significantly larger than zero only between 0.4 mHz and 10 mHz.
The blue line in the top panel represents the case in which the amplitude of both the astrophysical backgrounds is very constrained by external information.
This case reduces to the standard filter, as mentioned in section~\ref{s:new_filt}.
When $\sugb$ increases to 0.001 (orange line) the filter gives less weight to the frequencies around 1 mHz in order to reduce the response to the UGB signal --- which peaks at these frequencies, see the figure on the left.
When $\sugb$ reaches 0.01 (green line) the amplitude of the signal from galactic binaries is so uncertain that the filter is required to have zero response to the UGB signal shape.
This can only be achieved with a negative region in the filter: the trough around 1 mHz that the green lines have in all the panels.

A completely analogous discussion can be done about the BBH and BNS signal, whose amplitude with respect to the other components increases with frequency --- even though it is shadowed by the noise above 10 mHz.
This is the reason why the filter with $\segb = 0.01$ (middle panel) gives slightly less weight compared to the case with $\segb = 0.001$ (top panel) and the same region becomes negative for $\segb = 0.1$.

\begin{figure}
    \centering
    \includegraphics[width=0.49\linewidth]{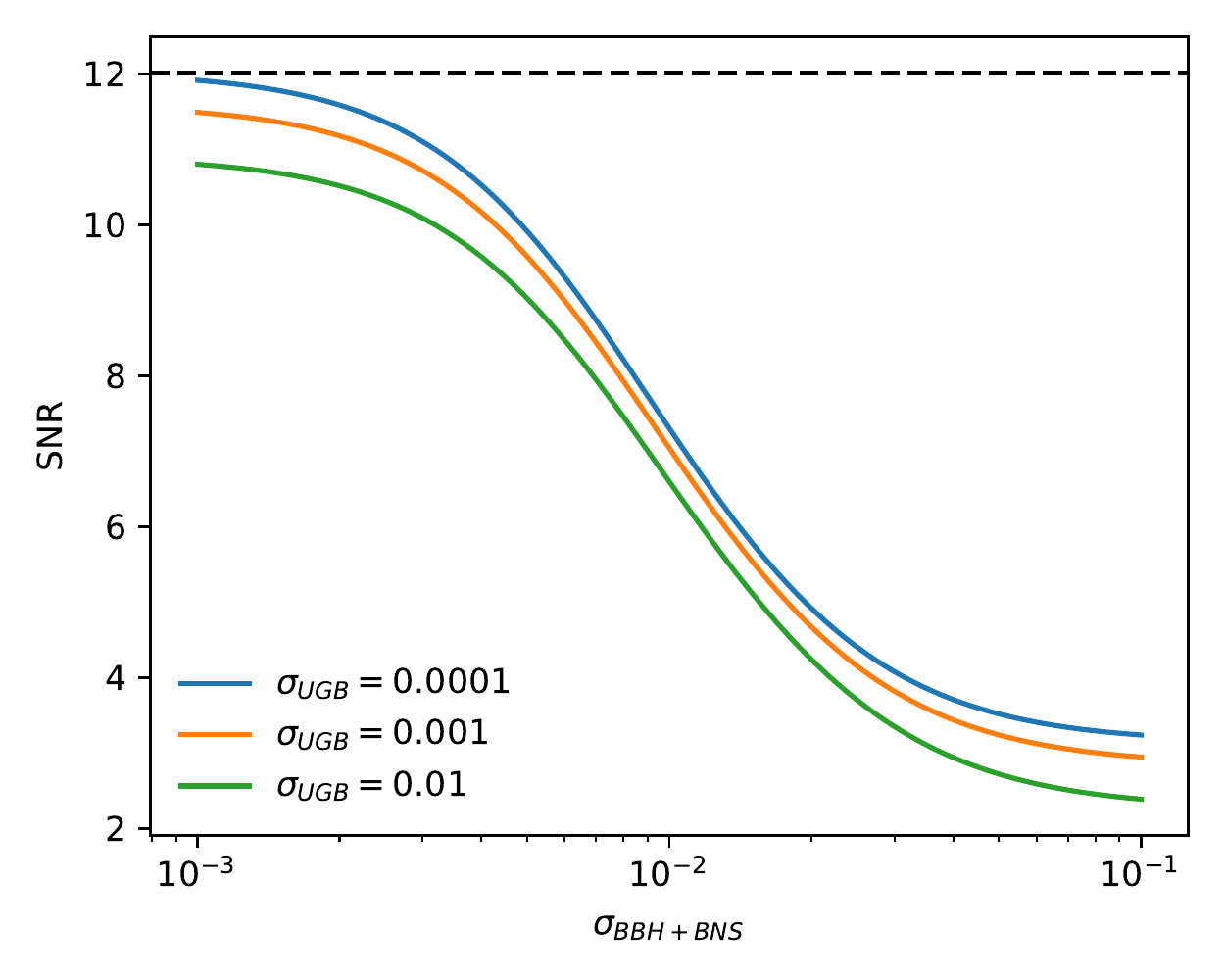}
    \includegraphics[width=0.49\linewidth]{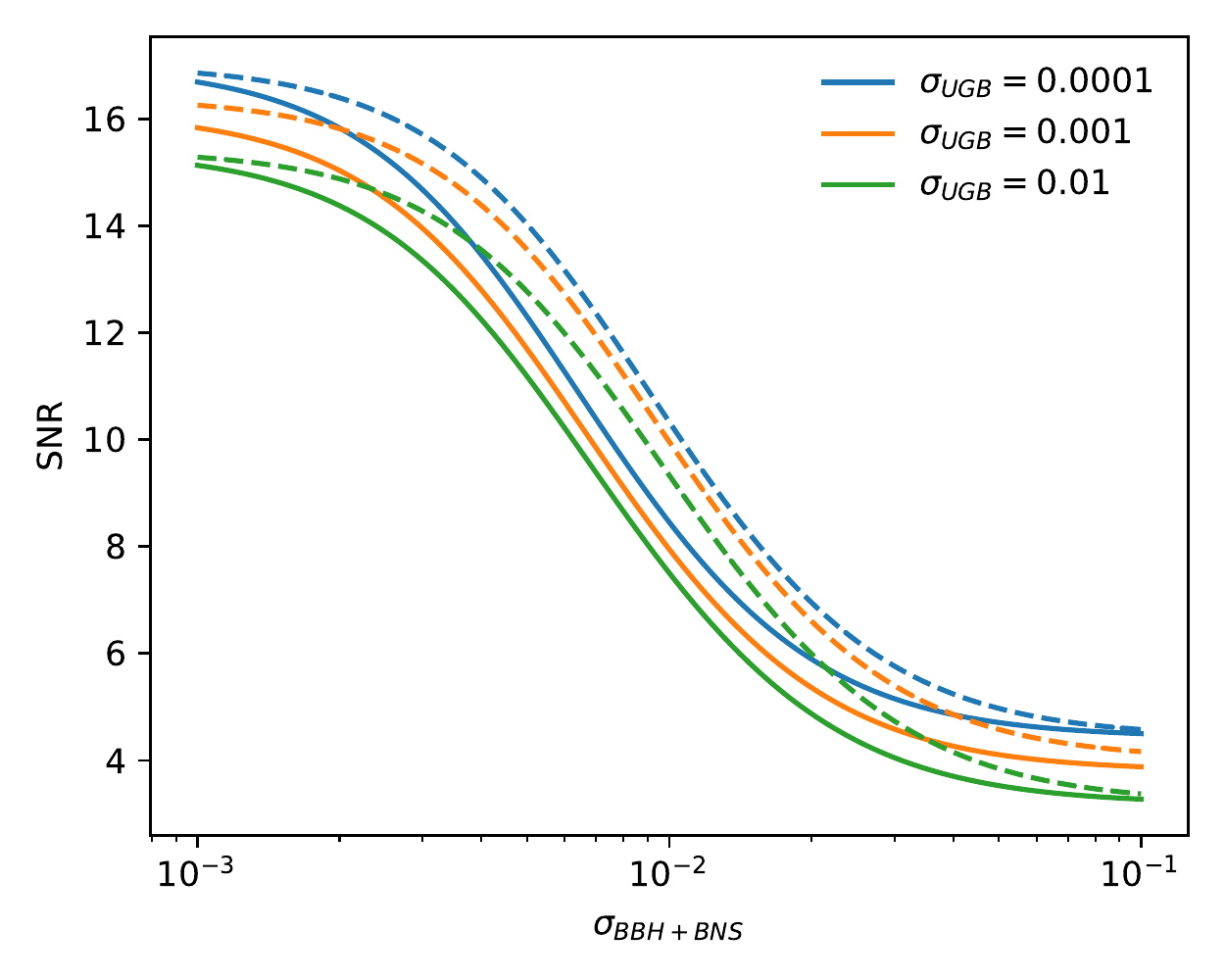}
    \caption{\emph{Left:} signal-to-noise ratio from either the $A$ or $E$ channel, for varying values of $\sigma_{\rm UGB}$ and $\sigma_{\rm BBH+BNS}$, computed with eq.~(\ref{e:new_snr})---actually eq.~(\ref{e:new_snr_vec}), because we are dealing with multiple astrophysical sources.
    For reference, the black dashed line reports the SNR obtained with the standard formula eq.~(\ref{e:old_snr}) ignoring the astrophysical components.
    \emph{Right:} signal-to-noise ratio from the $A$ and $E$ channels combined.
    The dashed line reports what one would get with a naive co-addition of the SNR of the individual channels (i.e.\ multiplying them by~$\sqrt{2}$).\label{f:snr}}
\end{figure}

As a final remark, we note that the peak of the filter increases when the amount of external information is reduced.
This is a consequence of the fact that when displaying these filters we impose that they have the same response to the target primordial signal.\footnote{We remind that the normalization of the filter is arbitrary and does not affect its~optimality.}
Therefore, the decreased (or negative) weight in the frequencies dominated by the astrophysical emission is compensated with an increased weight around the minimum of the astrophysical emission.
\paragraph{The SNR.}
We now study the SNR achieved for the same configuration and external information about the astrophysical components.
The result is reported in figure~\ref{f:snr}, where we also draw the standard SNR (black dashed line) computed with eq.~(\ref{e:old_snr}) ignoring the presence of astrophysical sources.
First note that this value coincides with the SNR obtained for $\segb = 0.0001$ and $\sugb= 0.001$, when the amplitude of the astrophysical signals is completely constrained by the external prior and therefore there is no information loss while marginalizing over them (blue line, left end).
Looking at the other extreme, if the amplitude of the astrophysical components is completely unconstrained ($\segb = 0.01$ and $\sugb = 0.1$, green line, right end), the rejection of their signal severely degrades the SNR from 12 to 2.4, which corresponds to a 96\% information loss.
This should be largely ascribed to the BBH+BNS, as even with $\segb = 0.0001$ (blue line) the information loss is still 93\%. The reason is simple: the frequency dependence of the primordial signal is much more orthogonal to the one of the UGB than the one of the BBH+BNS.

\setcounter{footnote}{3}

\subsection{Joint constraints from the A and E channels}
\label{s:application_template_free}
The only purpose of this section is to stress a supposedly obvious fact that is however easy to forget: using the same prior information across multiple measurements correlates the constraints they produce, and they can not be combined as if they were independent.

One of the most attractive properties of working with the AET channels compared to the XYZ channels is the fact that the cross-channel power is zero and they are statistically equivalent to independent experiments, at least in our idealized treatment.
Moreover, the A and E channels have the same noise level and, therefore, combining them effectively doubles the statistical information provided by the individual channels and, in the standard case eq.~(\ref{e:old_snr}), ${\rm SNR}^2_{A+E} = {\rm SNR}^2_{A} + {\rm SNR}^2_{E} = 2 {\rm SNR}^2_{A} $.

The formalism in section~\ref{s:new_filt}, however, can accommodate external information, which may play a significant role in ${\rm SNR}_A$ and ${\rm SNR}_E$ computed with eq.~(\ref{e:new_snr}).
When this happens the two SNRs are correlated and can not be added in quadrature.
Instead of accounting for the correlation in their coaddition, it is easier to properly compute the joint SNR\@.
It can be shown easily that the optimal combination of the two channels boils down to the average of the two auto-correlations, which has the same expected value and half the variance of the individual channels and is, therefore, equivalent to the $A$ (or $E$) channel with a $N$ divided by $\sqrt{2}$.
${\rm SNR}^2_{A+E}$ can be computed with the eq.~(\ref{e:new_snr}) applied to $A$ (or $E$) but multiplying every integral by 2.
Multiplying ${\rm SNR}^2_A$ by two altogether doubles the external information $\sigma^{-2}$ --- which, on the contrary, stays constant in ${\rm SNR}^2_{A+E}$.

In the right panel of figure~\ref{f:snr} we compare ${\rm SNR}_{A+E}$ with $\sqrt{2}\ {\rm SNR}_A$. The two coincide whenever the $\sigma$s are either very low or very high (the endpoints of the blue and green lines). When a $\sigma$ is very low, the correction term in eq.~(\ref{e:new_snr}) is negligible in both cases, while when the $\sigma$ is very high the correction term is not affected by its exact value.

Summarizing, the naive co-addition of the SNRs from independent measurements is significantly incorrect when the external information plays a significant but not overwhelming role. Note that this example of combination of the $A$ and $E$ channel is completely analogous to the combination of separate frequency bins. Naively co-adding the SNR from arbitrarily small bins produces a total SNR arbitrarily close to the standard SNR with no contaminants --- which is of course unphysical (and wrong).

\subsection{Template-free reconstruction with the A channel}
Without assuming prior knowledge on the primordial SGWB, we perform its reconstruction from simulated data of the $A$ channel following the prescription of section~\ref{s:template_free} (and appendix~\ref{s:fisher}).
We simulate $z_{AA}$ at linearly spaced frequencies, emulating what we would get if we were to compute it from the discrete Fourier transform of $d_A$.
The signal in $z_{AA}$ is the same primordial SGWB of the previous sections.
On the top of it we add Gaussian noise generated according to the covariance matrix in eq.~(\ref{e:z_var}), assuming $\sugb = 0.001$ and $\segb = 0.01$.
Note that this means that we are not only simulating independent realization of the instrumental noise, but also independent realizations of our prior distribution on the amplitude of the astrophysical components.
We create and analyze $10^4$ realizations.

The reconstruction is done using 8 non-overlapping top-hat (i.e., rectangular) functions with equal logarithmic width, covering the range from 0.1 mHz to 0.1 Hz.
This choice produces a result similar to binning $z_{AA}$ in logarithmic frequency intervals.
As explained in section~\ref{s:template_free}, more sophisticated sets of functions are possible, of course, but this simple choice already allows to show some of the main features of our approach.
Our estimator projects $z_{AA}$ onto the basis functions using the inverse variance in eq.~(\ref{e:fisher}). Therefore, the estimator not only does an inverse-noise-weighted average inside the bins, it also down-weights the spectral shapes that match the ones of the astrophysical signals --- the larger $\sigma$, the stronger the down-weighting.
This second effect is what distinguishes our estimator from a simple binning in this example.
To better highlight its role, we consider two estimators that differ only by the assumed value of $\sugb$ and $\segb$ and apply them to the same simulation set described above.
In one case we use the correct values of $\sugb$ and $\segb$ --- the same employed for simulating the data---, we label this case $\sigma \times 1$.
In a second case we assume one hundred times larger values and label it $\sigma \times 100$.
This latter case essentially coincides with a reconstruction that assumes no external information and will down-weight more aggressively the signals with the frequency dependence of the astrophysical components.

\begin{figure}
    \centering
    \includegraphics[width=\linewidth]{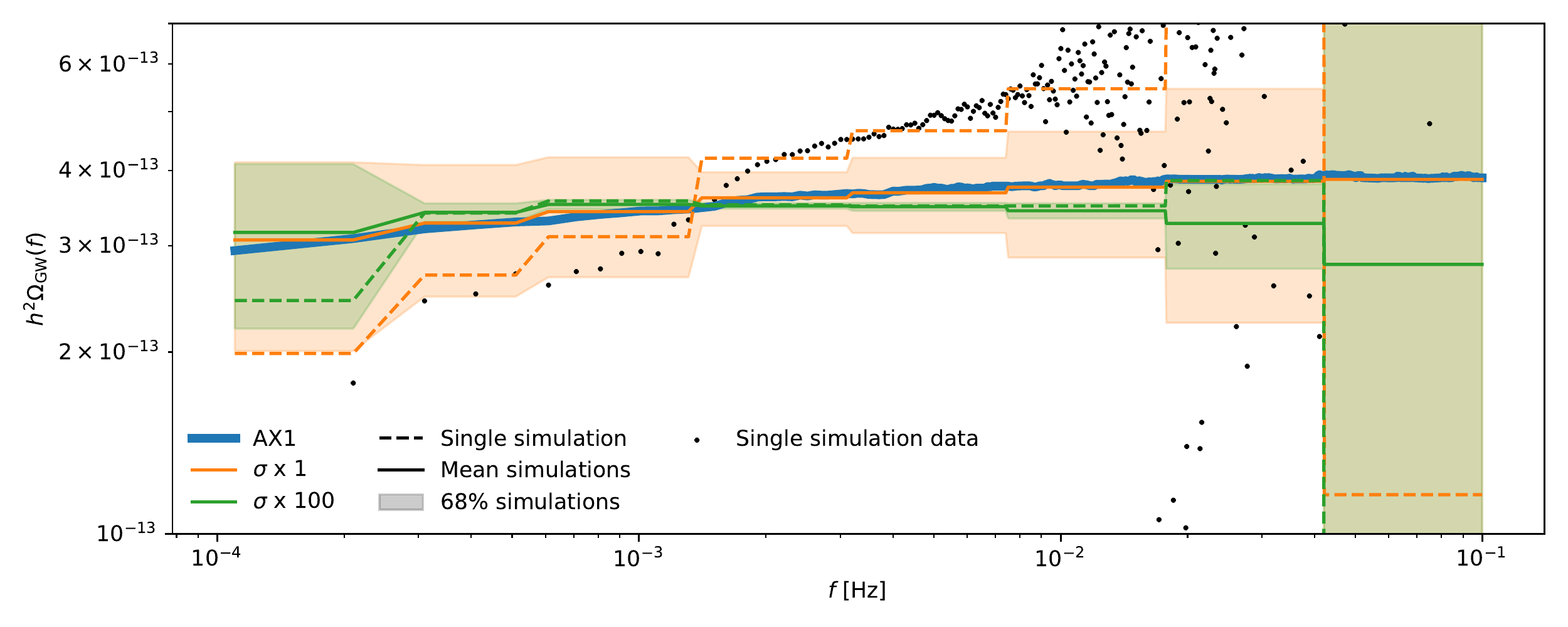}
    \caption{Template-free reconstruction. The thick blue line is the target signal.
    The orange represents the estimators that assume the same $\sugb$ and $\segb$ used for simulating the data, while the green assumes one hundred times larger values.
    The solid lines are the median of $10^4$ simulated reconstructions and the shaded areas cover from the 16$^{\rm th}$ to the 84$^{\rm th}$ percentile.
    As an example, we also show the data (dots) and the reconstruction (dashed lines) for a single simulation.\protect\footnotemark\label{f:template_free}}
\end{figure}

The results of the analysis are shown in figure~\ref{f:template_free}, which displays all the quantities ($z_{AA}$ included) in terms of energy density spectrum $\Omega_{\rm GW}$.
The figure reports both an example reconstruction of an individual simulation and a statistical summary of the reconstruction over the entire simulation set.
The $z_{AA}(f)$ of the individual simulation is represented by the dots, it clearly shows a considerable over-subtraction of the galactic signal and an under-subtraction of the extragalactic astrophysical signal.
The spectral shape of both signals is down-weighted by the estimators (dashed lines), which are indeed closer to the target signal than the original data.
In particular, the $\sigma \times 100$ estimator filters the spectral shapes of astrophysical origin more aggressively and, therefore, it seems to mitigate better the under- and over-subtraction of the astrophysical signals.
However, those modes are also removed from the target signal, and this is the reason why the median of the reconstructions (solid green line) is below the input primordial spectrum (blue line) above $\sim 2$ mHz.
Instead, the $\sigma \times 1$ estimator has a median very close to the target signal:
the external information, $\sigma^{-2}$, in the denominator of the second term in eq.~(\ref{e:fisher}) is a bit larger than the internal information and, therefore, that second term in eq.~(\ref{e:fisher}) slightly reduces the components of the primordial spectrum that resemble the ones of the astrophysical foregrounds, but not significantly.
Of course, the flip side of this is that if the data is substantially contaminated by the astrophysical foregrounds (like in the example shown in the figure) the estimation is closer to the data than to the target signal.

\footnotetext{To avoid making the figure too busy, we do not show the error bars of the single simulation, both for the data and the reconstructions. Those of the reconstructions would be equal to the boxes (and would be highly correlated, see the main text).}

The boxes in the figure~\ref{f:template_free} represent the standard deviation of the reconstructions across the entire simulation set and are linked to the uncertainty of the estimators.
For the $\sigma \times 1$ estimator (orange boxes) the scatter in the simulations is clearly dominated by the foreground contamination, while in the $\sigma \times 100$ case this component is filtered out and the dispersion is driven by the instrumental noise.
We want to stress, however, that this way of inferring uncertainties can be highly misleading: it represents only the diagonal of the covariance matrix, while the bins are highly correlated due to the foreground marginalization.
Once the full covariance is taken into account (see appendix~\ref{s:fisher}) one realizes that the ${\sigma \times 1}$ estimator is truly minimum variance, even if the modes that have the same frequency dependence of the astrophysical signals are highly uncertain.
They scatter less for the $\sigma \times 100$ estimator but just because their amplitude is systematically strongly suppressed and, therefore, their relative uncertainty actually tends to infinity.

\section{Discussion and conclusions}
\label{s:conclusions}
The existence of some form of SGWB is well motivated both from the theory and the direct GW observations of the past years.
The standard detection techniques have focused on distinguishing a template SGWB signal from the noise in the auto- or cross-correlation of interferometric signals.
From the data analysis point of view, there is freedom on the filter function involved in the correlation and the optimal choice is equal to the template signal divided by the noise power variance --- both diagonal in the frequency domain.

We extend this approach to the case in which other sources are present in addition to the target SGWB\@.
We derive the filter that optimally balances between the inverse-noise weighting done by the standard filter and the marginalization over the unwanted components, taking into account possible external priors.
Both the filter and the corresponding SNR are almost as simple as the standard ones --- and reduce to them either if the amount of external information is very large, or if the spectral shape of the unwanted components is very different from the target SGWB\@.
In particular, the optimal filter and SNR have closed forms and all the terms are easy to compute and interpret.
It is extremely simple, for example, to forecast the sensitivity of an experimental configuration to a primordial SGWB produced by an exotic inflationary model while, at the same time, marginalizing over a background $\Omega_{\rm GW} \propto f^{2/3}$ produced by unresolved binary black holes and neutron stars~\cite[see][]{Campeti:2020xwn}.
We apply the formalism to the LISA mission and show that neglecting the marginalization over the astrophysical signals --- as done by the standard estimator --- can grossly overestimate the SNR\@.
We remind that in the text we have used the terms ``primordial'' or ``astrophysical SGWB'' only for illustration purposes.
The approach is perfectly applicable to contexts where the target signal has an astrophysical origin or where both the target and unwanted component(s) are~primordial.

Our methodology is derived and applied in the context of isotropic, unpolarized background signals, but it can be easily generalized to polarized and anisotropic signals. For example, the analysis done by~\cite{Orlando:2020oko} can be amended to account for multiple sources of $V$. The only important caveat is that this methodology is based on the assumption (standard in the literature) that the variance in the auto- or cross-correlation is dominated by the noise at all frequencies. This assumption is key in obtaining nice and simple closed-form for the optimal filter and SNR, but may be too stringent in very high SNR settings.

Compared to techniques that blindly attempt to reconstruct the SGWB, the new filter we have illustrated does require a template of the target signal.
However, the filter can be relevant also to procedures that perform a more flexible SGWB reconstruction.
Take for example~\cite{Flauger:2020qyi}, the authors perform a template-free reconstruction of a primordial SGWB while marginalizing over a set of templates with uncertain amplitude (representing the same astrophysical components we considered in section~\ref{s:application}).
Their MCMC samples both the parameters of their non-linear primordial SGWB model and the amplitude of the astrophysical templates.
If the interest lies in the former and the latter are only marginalized over, this marginalization can be done analytically during the MCMC using the filter in eq.~(\ref{e:new_q}) (or analogous expressions) with a signal defined by the current value of the target signal parameters in the MCMC chain.
This allows to achieve the same result with reduced computational cost, as the parameters related to the astrophysical background are not explored by the Markov chain.

Moreover, we have shown how our methodology is applicable also to a template-free reconstruction of the SGWB\@.
Provided a basis for the admissible models, the method projects the data onto the space spanned by these models --- balancing noise-weighting and foreground removal.
The approach is similar to~\cite{Parida:2015fma}, the main difference is that we avoid estimating the amplitude of the astrophysical components by amending the frequency-frequency covariance matrix.\footnote{Another difference is that we allow for priors on the components we marginalize~over.}
We apply the methodology on a simulated LISA observation.
In our example we consider simple top-hat (rectangular) functions as the basis of the admissible models, but any choice is admitted --- such as polynomials, harmonic functions, or a basis of the possible SGWB produced by a class of inflationary models.
Albeit much richer than a single template, this type of procedure allows only linear models, but this class of parametrizations may turn out to be sufficiently flexible for many applications and fit the (simulated) data with an accuracy comparable to a more involved non-linear fit.

\acknowledgments

The author is thankful to Paolo Campeti, Eiichiro Komatsu, Carlo Baccigalupi, Mauro Pieroni and Davide Racco for useful comments and discussions.
The author acknowledges support from the ASI-COSMOS network (\url{www.cosmosnet.it}) and the INDARK INFN Initiative (\url{web.infn.it/CSN4/IS/Linea5/InDark}).

\appendix
\section{Derivation of the new filter}
\label{s:new_filt_dem}
We assume the presence of a primordial SGWB and a set of astrophysical backgrounds, due to different populations of unresolved sources,
\begin{equation}
    S_h(f) = S_p(f) + \sum_i \alpha_i S_{a_i}(f).
\end{equation}
We optimize the filter for the primordial SGWB but make no use of its properties in the derivation. Likewise, the astrophysical origin of the unwanted components plays no role.
The only relevant assumption for what follows is the fact that all the contributions have known frequency dependence and possibly unknown amplitude.\footnote{If useful, a free parameter can be introduced also in front of the primordial contribution. Nothing would change in the expression of the optimal filter and its derivation. The only difference would be the multiplication of the SNR by the expected value of such additional~parameter.}
For example, all the models that predict an $f^{-3/2}$ dependence can be accommodated (or approximated) in this formalism.

In the notation introduced in section~\ref{s:std_filt}, the signal can be written as
\begin{equation}
    \bm{h} = \bm{p} + \bm{A  \alpha},
\end{equation}
where the vector $\bm{p}$ is defined similarly to eq.~(\ref{e:vec_h}), with $S_p$ in place of $S_h$.
The same is done for the astrophysical sources, which are further collected in the columns of the matrix $\bm{A}$.
Their amplitudes (the coefficients $\alpha_i$) are free parameters and are collected in a vector $\bm{\alpha}$, with expected value $\bar{\bm{\alpha}}$ and covariance $\bm{\Sigma}$.
$\bm{A\alpha}$ and similar expressions represent standard linear algebra operations (matrix-vector product in the case).

We are interested in optimizing the filter for the detection of $\bm{p}$.
The cross-correlation can be corrected for the contribution of the astrophysical SGWB,
\begin{equation}
    y_{IJ} = x_{IJ} - \bm{q} \cdot \bm{A  \bar{\alpha}},
\end{equation}
so that its expected value depends only on the primordial SGWB,
\begin{equation}
    \langle y_{IJ} \rangle = \bm{q} \cdot \bm{p}.
\end{equation}
The variance is
\begin{equation}
    \langle y_{IJ}^2 \rangle - \langle y_{IJ} \rangle^2 =
    \bm{q}^t \bm{q}
    +
    \bm{q}^t \bm{A}\ \bm{\Sigma}\ \bm{A}^t \bm{q},
\end{equation}
where $\bm{a}^t \bm{b} \equiv \bm{a} \cdot \bm{b}$.
The new expression for the SNR is therefore
\begin{equation}
    {\rm SNR}^2 = \frac{(\bm{q}^t \bm{p})^2}{
    \bm{q}^t \bm{q}
    +
    \bm{q}^t \bm{A}\ \bm{\Sigma}\ \bm{A}^t \bm{q}}.
\end{equation}
Also in this case, the SNR is independent of the normalization of $\bm{q}$. We choose it such that
\begin{equation}
\bm{q}^t \bm{p} = \gamma,
\label{e:norm_q}
\end{equation}
where $\gamma$ is an arbitrary constant.
We now minimize the denominator of eq.~(\ref{e:new_snr}) under the condition in eq.~(\ref{e:norm_q}), which can be readily done using Lagrange multipliers.
The gradient of the Lagrangian is
\begin{equation}
\nabla_{\bm{q}} \mathcal{L}(\bm{q}, \lambda) =
2 (\bm{1} + \bm{A}\ \bm{\Sigma}\ \bm{A}^t) \bm{q} + \lambda \bm{p}.
\label{e:diff_L}
\end{equation}
Since the matrix in parenthesis is positive-defined, there is only the following stationary point and this point is a minimum
\begin{equation}
\bm{q} \propto (\bm{1} + \bm{A}\ \bm{\Sigma}\ \bm{A}^t)^{-1} \bm{p}.
\label{e:min_L}
\end{equation}
This expression is practically inconvenient because it involves the inversion of a large and dense matrix. Therefore, we use the Woodbury identity to re-express eq.~(\ref{e:min_L}) as
\begin{equation}
\bm{q} \propto  \bm{p} - \bm{A}\ (\bm{\Sigma}^{-1} + \bm{A}^t \bm{A})^{-1} \bm{A}^t  \bm{p},
\label{e:new_q_vec}
\end{equation}
which is the final form of our filter. The corresponding signal-to-noise ratio is
\begin{equation}
{\rm SNR}^2 = \bm{p}^t\bm{p} - \bm{p}^t\bm{A}\ (\bm{\Sigma}^{-1} + \bm{A}^t \bm{A})^{-1} \bm{A}^t  \bm{p}.
\label{e:new_snr_vec}
\end{equation}
For the case of one single astrophysical component, we report the explicit expression of this filter and SNR in eqs.~(\ref{e:new_q}) and~(\ref{e:new_snr}).

\section{Non-linear SGWB models}
\label{s:non_lin_model}
Many models of both cosmological and astrophysical SGWB contain non-linear parameters. Therefore, we generalize our model as follows
\begin{equation}
    S_h(f) = S_p(f) + \sum_i \alpha_i S_{a_i}(f, \beta_i),
\end{equation}
where $\beta_i$ is a non-linear parameter\footnote{It can also be a vector of parameters, the generalization is~straightforward.} in the model of the $i$-th astrophysical component.
By Taylor expanding the data model to first order
\begin{equation}
    S_h(f) \simeq S_p(f) + \sum_i \alpha_i S_{a_i}(f, \bar{\beta_i})  + \sum_i \bar{\alpha}_i \frac{\partial S_{a_i}}{\partial \beta_i}(f, \bar{\beta_i})\left(\beta_i - \bar{\beta}_i\right),
    \label{e:taylor}
\end{equation}
it becomes linear in the free parameters and therefore compatible with the assumptions made in the appendix~\ref{s:new_filt_dem}: $\bm{A}$ acquires extra columns given by all the $\bar{\alpha}_i \frac{\partial S_{a_i}}{\partial \beta_i}(f, \bar{\beta_i})$, and $\bm{\alpha}$ acquires extra rows given by all the $\beta_i - \bar{\beta}_i$ coefficients.
Of course, $\bm{\Sigma}$ acquires the same number of extra rows and columns, that can be used to accommodate external information about the non-linear parameters as well as their correlation with the linear ones.
Note that, if they are infinity (i.e.\ no external information is available on the non-linear parameters), the  $\bar{\alpha}_i$ factor multiplying the $i$-th new column of $\bm{A}$ is irrelevant: only the frequency dependence of the new column matters, not its normalization.

This natural extension of the formalism to non-linear parameters should, however, be used with care when analyzing data (or simulations).
First, one should consider if the model is overly complex for the experimental configuration being analyzed.
If the uncertainty on the linear parameters $\alpha_i$ is already very large, there is no point in refining the model of the $i$-th foreground.
Second, even if the constraints on those parameters are good, those on the non-linear parameters might be loose. As a result,
$\alpha_i S_{a_i}(f, \bar{\beta_i})  + \bar{\alpha}_i \frac{\partial S_{a_i}}{\partial \beta_i}(f, \bar{\beta_i})\left(\beta_i - \bar{\beta}_i\right)$
for the best fit $\alpha_i$ and $\beta_i$ may be quite far from what one would get form the fully non-linear fit of $\alpha_i S_{a_i}(f, \beta_i)$ --- meaning that the linear Taylor expansion eq.~(\ref{e:taylor}) is insufficient to describe the behaviour of the $i$-th astrophysical component in the range of plausible noise realizations.
This occurrence may not be easy to know in advance, it could even depend on the specific noise realization.
This may or may not be a problem for the constraints on the primordial component of the SGWB, depending on how degenerate $S_p(f)$ is with the difference between the linearized and fully non-linear fit.
The degeneracy is typically unknown from the onset and its evaluation would require computing the non-linear fit, which would make the linear approximation of limited interest.
Third, also the high signal-to-noise has a caveat: it still requires to guess $\bar{\beta_i}$.
If $\frac{\partial S_{a_i}}{\partial \beta_i}$ varies noticeably between $\bar{\beta_i}$ and the true value of $\beta_i$, the linearized and non-linear best-fit could be significantly different and bias the constraints on the primordial SGWB.

On the other hand, the applicability range of the linearized data model is much wider in the realm of forecasting.
The biases related to the noise realization and the imperfect choice of the reference $\bar{\beta_i}$ are indeed under control.
Moreover, if the quantity of interest is only the SNR, the value obtained with the linearized data model coincides with the Fisher estimate of the constraints from the non-linear fit.

\enlargethispage*{\baselineskip}\relax

\section{A closer look at the template-free reconstruction}
\label{s:fisher}
We start from the estimator in eq.~(\ref{e:z}), amended to account for an arbitrary number of astrophysical components
\begin{equation}
    z_{IJ}(f) \equiv \frac{2}{T \mathcal{R}_{IJ}(f)} \int_0^{\infty}d f' \left(d_I^*(f)d_J(f') + d_I(f)d_J^*(f')\right) \delta_T(f-f') - \frac{N_{IJ}(f)}{\mathcal{R}_{IJ}(f)} - \sum_i \bar{\alpha_i} S_{a_i}(f).
\end{equation}
Its covariance is equal to\footnote{It might be useful to remember that
\begin{equation*}
\begin{split}
\langle n^*_I(f)n_J(k)n_I(f')n^*_J(k') \rangle - \langle n^*_I(f)n_J(k)\rangle \langle n_I(f')n^*_J(k') \rangle = \frac{1}{4}[
&\delta( f- f') \delta( k- k') N_{II}(f) N_{JJ}(k) \\
&+ \delta( f+ k') \delta( f' + k) N_{IJ}(f) N_{IJ}(f') ]\,.
\end{split}
\end{equation*}}
\begin{equation}
    V(f, f') = \frac{N^2(f)\delta(f - f')}{2 T |\mathcal{R}_{IJ}(f)|^2} + \sum_i \sigma_{i}^2\ S_{a_i}(f)  S_{a_i}(f').
    \label{e:full_var}
\end{equation}
Since $\langle z_{IJ}(f)\rangle = S_p(f) $, $V$ is the frequency-frequency covariance for single-sided primordial signals.
The first term is the instrumental noise, while the second is the variance due to the fact that the normalization of the astrophysical component $a_i$ is known up to a standard deviation $\sigma_i$.

\begin{sloppypar}

To obtain readable expressions in what follows, we define $\bm{\Sigma} \equiv {\rm diag }(\sigma_i)$ and ${\bm{N} \equiv N^2(f)\delta(f - f')/2 T|\mathcal{R}_{IJ}(f)|^2}$  and collect all the astrophysical signals $S_{a_i}$ in the columns of $\bm{A}$. Note that the operator $\bm{A}$ is not exactly the same of appendix~\ref{s:new_filt_dem}: the basis of the frequency domain is different. We will also use a different (and more standard) inner product: $\bm{a}\cdot \bm{b} \equiv \int_0^\infty df a(f) b(f)$.
In this notation
\begin{equation}
    V(f, f') = \bm{N} + \bm{A} \bm{\Sigma} \bm{A}^t
\end{equation}
and its inverse can be obtained using the Woodbury identity
\begin{equation}
    F(f, f') = \bm{N}^{-1} - \bm{N}^{-1} \bm{A} \left( \bm{\Sigma}^{-1} + \bm{A}^t \bm{N}^{-1} \bm{A}\right)^{-1}\bm{A}^t \bm{N}^{-1}.
    \label{e:fisher_vec}
\end{equation}
This expression reduces to the eq.~(\ref{e:fisher}) in presence of only a single astrophysical source.
Note that, in spite of being very large, this matrix is easy to handle because it is a diagonal matrix plus a low-rank correction.
When no external information is available ($\bm{\Sigma}^{-1} = 0$), the term in parenthesis has the only role of making the columns of the $\bm{A}$ matrix $\bm{N}^{-1}$-orthonormal, so that the spectrum of $F$ on the space spanned by the foregrounds is zero (i.e., there is no information on any linear combination of the columns of $\bm{A}$).

\end{sloppypar}

Using $F$ as the inverse variance, we can also compute $\int_0^\infty df \int_0^\infty df' F(f, f')  z_{IJ}(f') z_{IJ}(f)$ to try to detect the presence of a signal in $z_{IJ}$ that can not be explained as a statistical fluctuation of the noise or the astrophysical signal.
Unfortunately, this can hardly be useful because this quantity is dominated by the immense number of low-SNR modes, which would likely shadow even fairly visible signals.
We can select a family of modes $\{m_i(f)\}_{i = 1, \dots, k}$ that we wish to use as a basis for the $k$-dimensional sub-space of admissible primordial backgrounds.
These modes can be used to probe the SNR on them one-by-one with eq.~(\ref{e:snr_info}).
It is probably more interesting, however, to project $z_{IJ}$ on the entire subspace generated by the models of our choice and to compute the total SNR\@.
Collecting the basis vectors in the columns of the matrix $\bm{M}$, the minimum-variance fit to the data is
\begin{equation}
    \hat{S}_p(f) = \bm{M} \left(\bm{M}^t \bm{F} \bm{M}\right)^{-1}\bm{M}^t \bm{F} \bm{z},
    \label{e:best_fit}
\end{equation}
and the SNR achieved is
\begin{equation}
    {\rm SNR}^2 = \bm{z}^t \bm{F}  \bm{M} \left(\bm{M}^t \bm{F} \bm{M}\right)^{-1}\bm{M}^t \bm{F} \bm{z}.
    \label{e:best_fit_snr}
\end{equation}
In the null hypothesis of no primordial signal present in the data, this SNR$^2$ is $\chi^2$-distributed with a number of degrees of freedom equal to $k$.
These are the familiar expressions produced by the generalized least squared estimator, which arise when estimating the parameters of a linear model under a constraint of minimum variance.
Note that, in general, the reconstructed amplitudes of the modes $\{m_i(f)\}_{i = 1, \dots, k}$ --- estimated using eq.~(\ref{e:best_fit}) without the leading $\bm{M}$ --- are correlated:
their covariance is  $ \left(\bm{M}^t \bm{F} \bm{M}\right)^{-1}$.

The estimator of~\cite{Parida:2015fma} has in fact the same expression of eq.~(\ref{e:best_fit}). In their approach $\bm{F}$ is simply $\bm{N}^{-1}$ and the foreground templates are added to the set of admissible modes $\bm{M}$.
Besides the fact that we allow for the presence of a prior on the amplitude of some of the components, the two estimations yield the same result as they are mathematically the same thing.
For low numbers of astrophysical components, they are also equivalent from the computational point of view, so choosing one way or the other is mostly a matter of taste.
Nevertheless, there is a scenario in which our approach provides an advantage.
If there is a degeneracy within the astrophysical components, in our approach it would manifest itself in the inversion of $\bm{\Sigma}^{-1} + \bm{A}^t \bm{N}^{-1} \bm{A}$.
Once the inversion is regularized, eq.~(\ref{e:best_fit}) can be computed and the degeneracy does not create any numerical instability. Instead, if the astrophysical components are included in $\bm{M}$, the degeneracy would manifest itself in the inversion of $\bm{M}^t \bm{F} \bm{M}$.
Also this inversion can be regularized, but in this case it is not trivial to say if the estimation of the primordial SGWB was compromised or not.

As already said in section~\ref{s:template_free}, the choice of the modes $\{m_i(f)\}_{i = 1, \dots, k}$ is probably the most important aspect of the template-free reconstruction and any choice is admissible: from classes of theoretical models to generic smooth functions.
We also mention PCA as a possible agnostic approach to the problem but we will now explain why we believe that this is probably not the most interesting application of the PCA\@.
This technique consists in choosing as $\{m_i(f)\}_{i = 1, \dots, k}$ the eigenvectors of the $k$ largest eigenvalues of $\bm{F}$ or, equivalently, the $k$ smallest of $\bm{V}$.
However, both matrices are written as diagonal matrices plus a low-rank correction.
Without the latter term the eigenvectors would be all delta functions and using the PCA would be equivalent to selecting modes in the trough of the noise curve.
The addition of a low-rank correction does not change significantly the result, as the eigenvectors still consist of a sharp feature very localized in the frequency range.
The overall effect is that PCA only narrows the frequency interval of $z_{IJ}(f)$ --- and possibly filters out the foreground templates, if they are very noisy.

We conclude this appendix with some computational remarks that can be relevant in this context when the number of sampled frequencies $n$ is large and down-sampling is not a viable solution.
\paragraph{Applying the inverse covariance matrix.}
The analytical expression in eq.~(\ref{e:fisher_vec}) is reasonably simple to implement and handle (most important, it is diagonal with a low-rank correction, as the covariance matrix).
Still one may prefer to work only with the covariance matrix instead of its inverse, in order to avoid the computation of
$\left(\bm{\Sigma}^{-1} + \bm{A}^t \bm{N}^{-1} \bm{A}\right)^{-1}$.

Eqs.~(\ref{e:best_fit}) and~(\ref{e:best_fit_snr}) can be computed using only the ability to apply $\bm{V}$ on a vector, which takes only a few $O(n)$ operations.\footnote{Needless to say, there is no reason to store the dense version of $\bm{V}$. It is sufficient to store only a few vectors: one for the noise and one for each astrophysical~component.}
It is sufficient to compute the solution to the system $\bm{VX} = \bm{M}$ and using $\bm{X}$ in place of $\bm{FM}$.
If solved with the preconditioned conjugate gradient (PCG), the solution is achieved at most in a number of iterations equal to the number of astrophysical components, thus with few $O(n)$ operations.
The reason is that the CG converges to the solution in a number of iterations lower than the number of distinct eigenvalues of the system matrix.
Once preconditioned with the inverse of the first term in eq.~(\ref{e:full_var}), the variance becomes the identity plus a low-rank correction and, therefore, the number of distinct eigenvalues is equal, at most, to this rank plus one.
\paragraph{PCA.}
As we said already, we do not expect this technique to play a significant role in the context we studied, especially if the number of astrophysical components is low.
Nevertheless, we show here that it is numerically doable even if $n$ is very large (and $k \ll n$).
The explicit calculation of the eigenvectors and eigenvalues of $\bm{F}$ (or $\bm{V}$) takes $O(n^2)$ storage locations and up to $O(n^3)$ operations.
However, we are not interested in the full set of eigenvectors and eigenvalues of the matrix but just in the $k$ best constrained by our experiment.
The eigenvectors with the highest eigenvalues can be computed in $O(k)$ applications of the matrix --- plus other $O(k)$ vector-vector products --- using the Lanczos method.
The overall $O(k \times n)$ scaling of this PCA poses no threat for any sensible value of $n$ and $k$. Note, however, that this scaling is accurate only for $k \ll n$.
When $k$ is comparable with $n$, other $O(k^2)$ calculations internal to the Lanczos method start to be significant.
The total scaling is still better than $O(n^3)$ but if $n$ is small the prefactors can be significant. Even more important, the explicit solution to the eigenvalue problem internally uses matrix-matrix operations, compared to the vector-vector operations inside the application of our sparse inverse-covariance. The former run much faster than the latter on modern CPUs (and GPUs), so the time-to-completion of an iterative $O(k \times n)$ algorithm may result comparable (or even higher!) than a $O(n^3)$ algorithm for small values of $n$.


\begin{thebibliography}{99}

\bibitem{Maggiore:2000gv}
%
M.~Maggiore,
\emph{Stochastic backgrounds of gravitational waves}, {\emph{ICTP Lect.\ Notes Ser.} {\bfseries 3} (2001) 397}
[\grqc{0008027}]
[\inspire{EPRINT\%2Bgr-qc\%2F0008027}].

\bibitem{Christensen:2018iqi}
%
N.~Christensen,
\emph{Stochastic Gravitational Wave Backgrounds},
\href{https://doi.org/10.1088/1361-6633/aae6b5}
{\emph{Rept.\ Prog.\ Phys.} {\bfseries 82} (2019) 016903}
[\arXivid{1811.08797}]
[\inspire{EPRINT\%2BarXiv\%3A1811.08797}].

\bibitem{Caprini:2018mtu}
%
C.~Caprini and D.G.~Figueroa,
\emph{Cosmological Backgrounds of Gravitational Waves},
\href{https://doi.org/10.1088/1361-6382/aac608}
{\emph{Class.\ Quant.\ Grav.} {\bfseries 35} (2018) 163001}
[\arXivid{1801.04268}]
[\inspire{EPRINT\%2BarXiv\%3A1801.04268}].

\bibitem{Regimbau:2011rp}
%
T.~Regimbau,
\emph{The astrophysical gravitational wave stochastic background},
\href{https://doi.org/10.1088/1674-4527/11/4/001}
{\emph{Res.\ Astron.\ Astrophys.} {\bfseries 11} (2011) 369}
[\arXivid{1101.2762}]
[\inspire{EPRINT\%2BarXiv\%3A1101.2762}].

\bibitem{Abbott:2016nmj}
%
{\scshape LIGO Scientific} and {\scshape Virgo} collaborations,
\emph{GW151226: Observation of Gravitational Waves from a 22-Solar-Mass Binary Black Hole Coalescence},
\href{https://doi.org/10.1103/PhysRevLett.116.241103}
{\emph{Phys.\ Rev.\ Lett.} {\bfseries 116} (2016) 241103}
[\arXivid{1606.04855}]
[\inspire{EPRINT\%2BarXiv\%3A1606.04855}].

\bibitem{Abbott:2016blz}
%
{\scshape LIGO Scientific} and {\scshape Virgo} collaborations,
\emph{Observation of Gravitational Waves from a Binary Black Hole Merger},
\href{https://doi.org/10.1103/PhysRevLett.116.061102}
{\emph{Phys.\ Rev.\ Lett.} {\bfseries 116} (2016) 061102}
[\arXivid{1602.03837}]
[\inspire{EPRINT\%2BarXiv\%3A1602.03837}].

\bibitem{TheLIGOScientific:2016pea}
%
{\scshape LIGO Scientific} and {\scshape Virgo} collaborations,
\emph{Binary Black Hole Mergers in the first Advanced LIGO Observing Run},
\href{https://doi.org/10.1103/PhysRevX.6.041015}
{\emph{Phys.\ Rev.\ X} {\bfseries 6} (2016) 041015}
[\erratum{8}{2018}{039903}]
 [\arXivid{1606.04856}]
[\inspire{EPRINT\%2BarXiv\%3A1606.04856}].

\bibitem{Abbott:2017vtc}
%
{\scshape LIGO Scientific} and {\scshape VIRGO} collaborations,
\emph{GW170104: Observation of a 50-Solar-Mass Binary Black Hole Coalescence at Redshift 0.2},
\href{https://doi.org/10.1103/PhysRevLett.118.221101}
{\emph{Phys.\ Rev.\ Lett.} {\bfseries 118} (2017) 221101}
[\erratum{121}{2018}{129901}]
 [\arXivid{1706.01812}]
[\inspire{EPRINT\%2BarXiv\%3A1706.01812}].

\bibitem{Abbott:2017gyy}
%
{\scshape LIGO Scientific} and {\scshape Virgo} collaborations,
\emph{GW170608: Observation of a 19-solar-mass Binary Black Hole Coalescence},
\href{https://doi.org/10.3847/2041-8213/aa9f0c}
{\emph{Astrophys.\ J.\ Lett.} {\bfseries 851} (2017) L35}
[\arXivid{1711.05578}]
[\inspire{EPRINT\%2BarXiv\%3A1711.05578}].

\bibitem{Abbott:2017oio}
%
{\scshape LIGO Scientific} and {\scshape Virgo} collaborations,
\emph{GW170814: A Three-Detector Observation of Gravitational Waves from a Binary Black Hole Coalescence},
\href{https://doi.org/10.1103/PhysRevLett.119.141101}
{\emph{Phys.\ Rev.\ Lett.} {\bfseries 119} (2017) 141101}
[\arXivid{1709.09660}]
[\inspire{EPRINT\%2BarXiv\%3A1709.09660}].

\bibitem{TheLIGOScientific:2017qsa}
%
{\scshape LIGO Scientific} and {\scshape Virgo} collaborations,
\emph{GW170817: Observation of Gravitational Waves from a Binary Neutron Star Inspiral},
\href{https://doi.org/10.1103/PhysRevLett.119.161101}
{\emph{Phys.\ Rev.\ Lett.} {\bfseries 119} (2017) 161101}
[\arXivid{1710.05832}]
[\inspire{EPRINT\%2BarXiv\%3A1710.05832}].

\bibitem{Buonanno:2004tp}
%
A.~Buonanno, G.~Sigl, G.G.~Raffelt, H.-T.~Janka and E.~Muller,
\emph{Stochastic gravitational wave background from cosmological supernovae},
\href{https://doi.org/10.1103/PhysRevD.72.084001}
{\emph{Phys.\ Rev.\ D} {\bfseries 72} (2005) 084001}
[\astroph{0412277}]
[\inspire{EPRINT\%2Bastro-ph\%2F0412277}].

\bibitem{Yakunin:2010fn}
%
K.N.~Yakunin et~al.,
\emph{Gravitational Waves from Core Collapse Supernovae},
\href{https://doi.org/10.1088/0264-9381/27/19/194005}
{\emph{Class.\ Quant.\ Grav.} {\bfseries 27} (2010) 194005}
[\arXivid{1005.0779}]
[\inspire{EPRINT\%2BarXiv\%3A1005.0779}].

\bibitem{Ferrari:1998jf}
%
V.~Ferrari, S.~Matarrese and R.~Schneider,
\emph{Stochastic background of gravitational waves generated by a cosmological population of young, rapidly rotating neutron stars},
\href{https://doi.org/10.1046/j.1365-8711.1999.02207.x}
{\emph{Mon.\ Not.\ Roy.\ Astron.\ Soc.} {\bfseries 303} (1999) 258}
[\astroph{9806357}]
[\inspire{EPRINT\%2Bastro-ph\%2F9806357}].

\bibitem{Cheng:2015rja}
%
Q.~Cheng, Y.-W.~Yu and X.-P.~Zheng,
\emph{Stochastic gravitational wave background from magnetic deformation of newly born magnetars},
\href{https://doi.org/10.1093/mnras/stv2127}
{\emph{Mon.\ Not.\ Roy.\ Astron.\ Soc.} {\bfseries 454} (2015) 2299}
[\arXivid{1509.07651}]
[\inspire{EPRINT\%2BarXiv\%3A1509.07651}].

\bibitem{Starobinsky:1980te}
%
A.A.~Starobinsky,
\emph{A New Type of Isotropic Cosmological Models Without Singularity},
\href{https://doi.org/10.1016/0370-2693(80)90670-X}
{\emph{Adv. Ser. Astrophys. Cosmol.} {\bfseries 3} (1987) 130}
[\inspire{J\%20\%22Adv.Ser.Astrophys.Cosmol.\%2C3\%2C130\%22}].

\bibitem{Linde:1981mu}
%
A.D.~Linde,
\emph{A New Inflationary Universe Scenario: A Possible Solution of the Horizon, Flatness, Homogeneity, Isotropy and Primordial Monopole Problems},
\href{https://doi.org/10.1016/0370-2693(82)91219-9}
{\emph{Adv. Ser. Astrophys. Cosmol.} {\bfseries 3} (1987) 149}
[\inspire{J\%20\%22Adv.Ser.Astrophys.Cosmol.\%2C3\%2C149\%22}].

\bibitem{Guth:1980zm}
%
A.H.~Guth,
\emph{The Inflationary Universe: A Possible Solution to the Horizon and Flatness Problems},
\href{https://doi.org/10.1103/PhysRevD.23.347}
{\emph{Adv. Ser. Astrophys. Cosmol.} {\bfseries 3} (1987) 139}
[\inspire{J\%20\%22Adv.Ser.Astrophys.Cosmol.\%2C3\%2C139\%22}].

\bibitem{Mukhanov:1981xt}
%
V.F.~Mukhanov and G.V.~Chibisov,
\emph{Quantum Fluctuations and a Nonsingular Universe}, {\emph{JETP Lett.} {\bfseries 33} (1981) 532}
[\inspire{J\%20\%22JETP\%20Lett.\%2C33\%2C532\%22}].

\bibitem{Bartolo:2016ami}
%
N.~Bartolo et~al.,
\emph{Science with the space-based interferometer LISA. IV: Probing inflation with gravitational waves},
\jcap{12}{2016}{026}
[\arXivid{1610.06481}]
[\inspire{EPRINT\%2BarXiv\%3A1610.06481}].

\bibitem{Guzzetti:2016mkm}
%
M.C.~Guzzetti, N.~Bartolo, M.~Liguori and S.~Matarrese,
\emph{Gravitational waves from inflation},
\href{https://doi.org/10.1393/ncr/i2016-10127-1}
{\emph{Riv.\ Nuovo Cim.} {\bfseries 39} (2016) 399}
[\arXivid{1605.01615}]
[\inspire{EPRINT\%2BarXiv\%3A1605.01615}].

\bibitem{Hazumi:2019lys}
%
M.~Hazumi et~al.,
\emph{LiteBIRD: A Satellite for the Studies of B-Mode Polarization and Inflation from Cosmic Background Radiation Detection},
\href{https://doi.org/10.1007/s10909-019-02150-5}
{\emph{J.\ Low Temp.\ Phys.} {\bfseries 194} (2019) 443}
[\inspire{J\%20\%22J.Low\%20Temp.Phys.\%2C194\%2C443\%22}].

\bibitem{Abazajian:2019eic}
%
K.~Abazajian et~al.,
\emph{CMB-S4 Science Case, Reference Design, and Project Plan},
\arXivid{1907.04473}
[\inspire{EPRINT\%2BarXiv\%3A1907.04473}].

\bibitem{Kibble:1976sj}
%
T.W.B.~Kibble,
\emph{Topology of Cosmic Domains and Strings},
\href{https://doi.org/10.1088/0305-4470/9/8/029}
{\emph{J.\ Phys.\ A} {\bfseries 9} (1976) 1387}
[\inspire{J\%20\%22J.Phys.\%2CA9\%2C1387\%22}].

\bibitem{Vachaspati:2015cma}
%
T.~Vachaspati, L.~Pogosian and D.~Steer,
\emph{Cosmic Strings},
\href{https://doi.org/10.4249/scholarpedia.31682}
{\emph{Scholarpedia} {\bfseries 10} (2015) 31682}
[\arXivid{1506.04039}]
[\inspire{EPRINT\%2BarXiv\%3A1506.04039}].

\bibitem{Sarangi:2002yt}
%
S.~Sarangi and S.H.H.~Tye,
\emph{Cosmic string production towards the end of brane inflation},
\href{https://doi.org/10.1016/S0370-2693(02)01824-5}
{\emph{Phys.\ Lett.\ B} {\bfseries 536} (2002) 185}
[\hepth{0204074}]
[\inspire{EPRINT\%2Bhep-th\%2F0204074}].

\bibitem{Jones:2003da}
%
N.T.~Jones, H.~Stoica and S.H.H.~Tye,
\emph{The Production, spectrum and evolution of cosmic strings in brane inflation},
\href{https://doi.org/10.1016/S0370-2693(03)00592-6}
{\emph{Phys.\ Lett.\ B} {\bfseries 563} (2003) 6}
[\hepth{0303269}]
[\inspire{EPRINT\%2Bhep-th\%2F0303269}].

\bibitem{Pagano:2015hma}
%
L.~Pagano, L.~Salvati and A.~Melchiorri,
\emph{New constraints on primordial gravitational waves from Planck 2015},
\href{https://doi.org/10.1016/j.physletb.2016.07.078}
{\emph{Phys.\ Lett.\ B} {\bfseries 760} (2016) 823}
[\arXivid{1508.02393}]
[\inspire{EPRINT\%2BarXiv\%3A1508.02393}].

\bibitem{Abbott:2017mem}
%
{\scshape LIGO Scientific} and {\scshape Virgo} collaborations,
\emph{Constraints on cosmic strings using data from the first Advanced LIGO observing run},
\href{https://doi.org/10.1103/PhysRevD.97.102002}
{\emph{Phys.\ Rev.\ D} {\bfseries 97} (2018) 102002}
[\arXivid{1712.01168}]
[\inspire{EPRINT\%2BarXiv\%3A1712.01168}].

\bibitem{TheLIGOScientific:2016wyq}
%
{\scshape LIGO Scientific} and {\scshape Virgo} collaborations,
\emph{GW150914: Implications for the stochastic gravitational wave background from binary black holes},
\href{https://doi.org/10.1103/PhysRevLett.116.131102}
{\emph{Phys.\ Rev.\ Lett.} {\bfseries 116} (2016) 131102}
[\arXivid{1602.03847}]
[\inspire{EPRINT\%2BarXiv\%3A1602.03847}].

\bibitem{Abbott:2017xzg}
%
{\scshape LIGO Scientific} and {\scshape Virgo} collaborations,
\emph{GW170817: Implications for the Stochastic Gravitational-Wave Background from Compact Binary Coalescences},
\href{https://doi.org/10.1103/PhysRevLett.120.091101}
{\emph{Phys.\ Rev.\ Lett.} {\bfseries 120} (2018) 091101}
[\arXivid{1710.05837}]
[\inspire{EPRINT\%2BarXiv\%3A1710.05837}].

\bibitem{Arzoumanian:2020vkk}
%
{\scshape NANOGrav} collaboration,
\emph{The NANOGrav 12.5 yr Data Set: Search for an Isotropic Stochastic Gravitational-wave Background},
\href{https://doi.org/10.3847/2041-8213/abd401}
{\emph{Astrophys.\ J.\ Lett.} {\bfseries 905} (2020) L34}
[\arXivid{2009.04496}]
[\inspire{EPRINT\%2BarXiv\%3A2009.04496}].

\bibitem{Hellings:1983fr}
%
R.w.~Hellings and G.s.~Downs,
\emph{Upper limits on the isotropic gravitational radiation background from pulsar timing analysis},
\href{https://doi.org/10.1086/183954}
{\emph{Astrophys.\ J.\ Lett.} {\bfseries 265} (1983) L39}
[\inspire{J\%20\%22Astrophys.J.Lett.\%2C265\%2CL39\%22}].

\bibitem{Pan:2019uyn}
%
Z.~Pan and H.~Yang,
\emph{Probing Primordial Stochastic Gravitational Wave Background with Multi-band Astrophysical Foreground Cleaning},
\href{https://doi.org/10.1088/1361-6382/abb074}
{\emph{Class.\ Quant.\ Grav.} {\bfseries 37} (2020) 195020}
[\arXivid{1910.09637}]
[\inspire{EPRINT\%2BarXiv\%3A1910.09637}].

\bibitem{Flauger:2020qyi}
%
R.~Flauger, N.~Karnesis, G.~Nardini, M.~Pieroni, A.~Ricciardone and J.~Torrado,
\emph{Improved reconstruction of a stochastic gravitational wave background with LISA},
\jcap{01}{2021}{059}
[\arXivid{2009.11845}]
[\inspire{EPRINT\%2BarXiv\%3A2009.11845}].

\bibitem{Pieroni:2020rob}
%
M.~Pieroni and E.~Barausse,
\emph{Foreground cleaning and template-free stochastic background extraction for LISA},
\jcap{07}{2020}{021}
[\erratum{09}{2020}{E01}]
 [\arXivid{2004.01135}]
[\inspire{EPRINT\%2BarXiv\%3A2004.01135}].

\bibitem{Parida:2015fma}
%
A.~Parida, S.~Mitra and S.~Jhingan,
\emph{Component Separation of a Isotropic Gravitational Wave Background},
\jcap{04}{2016}{024}
[\arXivid{1510.07994}]
[\inspire{EPRINT\%2BarXiv\%3A1510.07994}].

\bibitem{Allen:1996vm}
%
B.~Allen,
\emph{The stochastic gravity wave background: Sources and detection}, in
\emph{{Les Houches School of Physics: Astrophysical Sources of Gravitational Radiation}}, (1996), pp.~373--417,
[\grqc{9604033}]
[\inspire{EPRINT\%2Bgr-qc\%2F9604033}].

\bibitem{Allen:1997ad}
%
B.~Allen and J.D.~Romano,
\emph{Detecting a stochastic background of gravitational radiation: Signal processing strategies and sensitivities},
\href{https://doi.org/10.1103/PhysRevD.59.102001}
{\emph{Phys.\ Rev.\ D} {\bfseries 59} (1999) 102001}
[\grqc{9710117}]
[\inspire{EPRINT\%2Bgr-qc\%2F9710117}].

\bibitem{2012CQGra..29l4015V}
%
M.~Vallisneri and C.R.~Galley,
\emph{Non-sky-averaged sensitivity curves for space-based gravitational-wave observatories},
\href{https://doi.org/10.1088/0264-9381/29/12/124015}
{\emph{Class.\ Quant.\ Grav.} {\bfseries 29} (2012) 124015}
[\arXivid{1201.3684}]
[\inspire{EPRINT\%2BarXiv\%3A1201.3684}].

\bibitem{2017arXiv170200786A}
%
{\scshape LISA} collaboration,
\emph{Laser Interferometer Space Antenna},
\arXivid{1702.00786}
[\inspire{EPRINT\%2BarXiv\%3A1702.00786}].

\bibitem{Smith:2019wny}
%
T.L.~Smith and R.~Caldwell,
\emph{LISA for Cosmologists: Calculating the Signal-to-Noise Ratio for Stochastic and Deterministic Sources},
\href{https://doi.org/10.1103/PhysRevD.100.104055}
{\emph{Phys.\ Rev.\ D} {\bfseries 100} (2019) 104055}
[\arXivid{1908.00546}]
[\inspire{EPRINT\%2BarXiv\%3A1908.00546}].

\bibitem{LIGOScientific:2019vic}
%
{\scshape LIGO Scientific} and {\scshape Virgo} collaborations,
\emph{Search for the isotropic stochastic background using data from Advanced LIGO's second observing run},
\href{https://doi.org/10.1103/PhysRevD.100.061101}
{\emph{Phys.\ Rev.\ D} {\bfseries 100} (2019) 061101}
[\arXivid{1903.02886}]
[\inspire{EPRINT\%2BarXiv\%3A1903.02886}].

\bibitem{Cornish:2017vip}
%
N.~Cornish and T.~Robson,
\emph{Galactic binary science with the new LISA design},
\href{https://doi.org/10.1088/1742-6596/840/1/012024}
{\emph{J.\ Phys.\ Conf.\ Ser.} {\bfseries 840} (2017) 012024}
[\arXivid{1703.09858}]
[\inspire{EPRINT\%2BarXiv\%3A1703.09858}].

\bibitem{Cornish:2018dyw}
%
T.~Robson, N.J.~Cornish and C.~Liu,
\emph{The construction and use of LISA sensitivity curves},
\href{https://doi.org/10.1088/1361-6382/ab1101}
{\emph{Class.\ Quant.\ Grav.} {\bfseries 36} (2019) 105011}
[\arXivid{1803.01944}]
[\inspire{EPRINT\%2BarXiv\%3A1803.01944}].

\bibitem{Schmitz:2020rag}
%
K.~Schmitz,
\emph{LISA Sensitivity to Gravitational Waves from Sound Waves},
\href{https://doi.org/10.3390/sym12091477}
{\emph{Symmetry} {\bfseries 12} (2020) 1477}
[\arXivid{2005.10789}]
[\inspire{EPRINT\%2BarXiv\%3A2005.10789}].

\bibitem{Campeti:2020xwn}
%
P.~Campeti, E.~Komatsu, D.~Poletti and C.~Baccigalupi,
\emph{Measuring the spectrum of primordial gravitational waves with CMB, PTA and Laser Interferometers},
\jcap{01}{2021}{012}
[\arXivid{2007.04241}]
[\inspire{EPRINT\%2BarXiv\%3A2007.04241}].

\bibitem{Adams:2013qma}
%
M.R.~Adams and N.J.~Cornish,
\emph{Detecting a Stochastic Gravitational Wave Background in the presence of a Galactic Foreground and Instrument Noise},
\href{https://doi.org/10.1103/PhysRevD.89.022001}
{\emph{Phys.\ Rev.\ D} {\bfseries 89} (2014) 022001}
[\arXivid{1307.4116}]
[\inspire{EPRINT\%2BarXiv\%3A1307.4116}].

\bibitem{Orlando:2020oko}
%
G.~Orlando, M.~Pieroni and A.~Ricciardone,
\emph{Measuring Parity Violation in the Stochastic Gravitational Wave Background with the LISA-Taiji network},
\jcap{03}{2021}{069}
[\arXivid{2011.07059}]
[\inspire{EPRINT\%2BarXiv\%3A2011.07059}].



\end{thebibliography}

\end{document}